\documentclass[10pt,aps,pra,amsfonts,amssymb,amsmath,
floatfix,nofootinbib, groupedaddress,superscriptaddress,citesort,twocolumn
]{revtex4}%

\usepackage{graphicx}
\usepackage{subfigure}       
\usepackage{caption}      
\usepackage{xcolor}       
\usepackage{amsthm}
\usepackage{mathrsfs}
\usepackage{amsfonts}
\usepackage{amstext}
\usepackage{bm}
\usepackage{bbm}
\usepackage{hyperref}     
\usepackage{caption} 
\makeatletter
\let\old@makecaption\@makecaption
\renewcommand{\@makecaption}[2]{\old@makecaption{#1}{\ignorespaces #2}}
\makeatother

\hypersetup{
    hyperindex=true,
    breaklinks=true,
    hidelinks,
    colorlinks=false
}

\newtheorem{theorem}{Theorem}

\newtheorem{lemma}[theorem]{Lemma}

\newtheorem{example}[theorem]{Example}
\newtheorem{definition}[theorem]{Definition}

\begin{document}
	
	\preprint{APS/123-QED}
	
	\title{An easily computable measure of Gaussian quantum imaginarity}
	
	\author{Ting Zhang}
	\affiliation{School of Mathematics and Statistics, Shanxi University,
		Taiyuan 030006, P. R. China}

	\author{Jinchuan Hou}
	\email{jinchuanhou@aliyun.com}\affiliation{College of Mathematics, Taiyuan University of
		Technology, Taiyuan 030024, P. R. China}
	
\author{Xiaofei Qi}
\email{xiaofeiqisxu@aliyun.com}
	\affiliation{School of Mathematics and Statistics, Shanxi University,
		Taiyuan 030006, P. R. China}\affiliation{Key Laboratory of Complex Systems and Data Science of Ministry of Education,
		Shanxi University, Taiyuan  030006,  Shanxi, China}

\begin{abstract}
The resource-theoretic frameworks for quantum imaginarity have been developed in recent years. Within these frameworks, many  imaginarity measures for finite-dimensional systems have been proposed. However, for imaginarity of Gaussian states in continuous-variable (CV) systems, there are only two known Gaussian imaginarity measures, which exhibit prohibitive computational complexity when applied to multi-mode Gaussian states. In this paper, we propose a computable Gaussian imaginarity measure $\mathcal I^{G_n}$ for $n$-mode Gaussian systems.  The value of $\mathcal I^{G_n}$ is simply formulated by the displacement vectors and covariance matrices of Gaussian states.  A comparative analysis of $\mathcal{I}^{G_n}$ with existing two Gaussian imaginarity measures indicates that $\mathcal{I}^{G_n}$ can be used to detect imaginarity in any $n$-mode Gaussian states more efficiently.  As an application, we study the dynamics behaviour of $(1+1)$-mode Gaussian states in Gaussian Markovian noise environments for two-mode CV system by utilizing  ${\mathcal I}^{G_2}$.
Moreover,  we prove that, ${\mathcal I}^{G_n}$ can induce a quantification of any  $m$-multipartite multi-mode CV systems which  satisfies all requirements for  measures of multipartite multi-mode Gaussian correlations, which unveils  that, $n$-mode Gaussian imaginarity can  also be regarded as a kind of multipatite multi-mode Gaussian correlation and is  a multipartite Gaussian quantum resource.

\end{abstract}
\maketitle

\section{Introduction}

Complex numbers are indispensable parts of describing quantum states and their dynamic behaviors, and  are widely used in classical and quantum physics.
Since the emergence of quantum mechanics, there has been an enduring scientific discourse why the theoretical framework of quantum mechanics
must use complex numbers, rather than being solely based on pure real numbers. Despite numerous proposals aiming to formulate quantum mechanics using real numbers,
these attempts have failed to fully capture the complexity and multifaceted nature of quantum phenomena.
The latest advancements in quantum resource theory provide a clear and definitive answer: imaginarity is a quantum resource \cite{WKR2,RTW,CWL,LMW}.
From then on, a surge of investigations emerged to focus  on the imaginarity theory. Research indicated that imaginarity, as a pivotal resource, plays a crucial role in various fields, such as  hiding and masking quantum information \cite{Zhu}, multiparameter metrology \cite{MM}, machine learning algorithms \cite{SSS},
Kirkwood-Dirac quasiprobability distributions \cite{Bu,BD,BAN,WSP}, weak-value theory \cite{WG} and the nonlocal advantages exhibited by quantum imaginarity \cite{WF}.

A general quantum resource theory  is composed of three fundamental elements: free states (those states having no resource), free operations (those quantum channels that send free states into free states) and measures nonincreasing under free operations.
In the resource theory of imaginarity, the free states are real states and the free operations are real operations \cite{HO,CG}. Concretely, assume that $H$ is a complex finite-dimensional Hilbert space and ${\mathcal S}(H)$ is the set of all quantum states on $H$. For any fixed reference basis ${|j\rangle}_{j=1}^d$ of $H$, a quantum state $\rho\in{\mathcal S}(H)$ is called a real state if $\langle j|\rho|k\rangle\in\mathbb{R}$ holds for all $j,k\in\{1,2,\cdots, d\}$; a quantum operation $\Phi:{\mathcal S}(H)\rightarrow{\mathcal S}(H)$ with Krauss representation $\Phi(\rho)=\sum_lK_l\rho K_l^{\dag}$ is called a real operation if its all Kraus operators are real, that is, $\langle j|K_l|k\rangle\in\mathbb{R}$ for all $l,j,k$ \cite{HG}.

Within the framework of imaginarity resource theory, an essential challenge lies in quantifying imaginarity. Formally, a nonnegative functional $\mathcal{I}:{\mathcal S}(H)\rightarrow \mathbb [0,+\infty)$ is  referred to be an imaginarity measure if it fulfills the following conditions (I1)-(I4) (see \cite{WKR1,WKR2,HG}):

(I1) Faithfulness:  $\mathcal{I}(\rho)=0$ if and only if $\rho$ is  real.

(I2) Monotonicity: $\mathcal{I}(\Phi(\rho))\le\mathcal{I}(\rho)$ holds for all real quantum operations $\Phi$ and all $\rho\in{\mathcal S}(H)$.

(I3) Probabilistic monotonicity:  $$\sum_l{\rm Tr}(K_l\rho K_l^\dag)\mathcal{I}(\frac{K_l\rho K_l^\dag}{{\rm Tr}(K_l\rho K_l^\dag)})\le\mathcal{I}(\rho)$$
for all $\rho\in{\mathcal S}(H)$, where  $\{K_l\}$ are real Kraus operators with $\sum_l K_l^\dag K_l=I$.

(I4) Convexity: $\mathcal{I}(\sum_jp_j\rho_j)\le\sum_jp_j\mathcal{I}(\rho_j)$ for any probability distribution $\{p_j\}$ and quantum states $\{\rho_j\}$.

It is worth noting that (I3) and (I4) can jointly imply (I2). In \cite{XGL}, the authors introduced the following condition.

(I5) Additivity for direct sum states:
$\mathcal{I}(p\rho_1\oplus(1-p)\rho_2)=p\mathcal{I}(\rho_1)+(1-p)\mathcal{I}(\rho_2),$ where $p\in(0,1)$, $\rho_1$ and $\rho_2$ are any quantum states.

It is shown in \cite{XGL} that (I1)-(I4) are equivalent to (I1), (I2) and (I5).
Several imaginarity measures in finite-dimensional quantum systems are proposed,  for instance, the trace norm and $l_1$-norm of imaginarity \cite{WKR2,HG,CGY}, the robustness of imaginarity \cite{WKR1}, geometric imaginarity \cite{WKR1,XXZ}, fidelity-based imaginarity \cite{WKR1,KDS}, the relative entropy of imaginarity \cite{XGL}, the concurrence measure of imaginarity \cite{WKS}, convex roof imaginarity and the least imaginarity of the input pure states under real operations \cite{DB}.

The continuous-variable (CV) quantum systems are  equally  important from both theoretical and experimental perspectives. Therefore it is also important and
interesting to study imaginarity in CV systems. In CV systems, there exists a kind of important quantum states, that is, Gaussian states, which  can be readily produced and manipulated in experimental arrangements, and can be  extensively utilized in quantum optics and quantum information theory \cite{CP,BV,WHT,Leo,WPG,Oli,ARL,Ser}.
The concept of real Gaussian states and real Gaussian operations were introduced and  characterized  in \cite{Xu1} recently. Recall that a real Gaussian operation is a Gaussian channel that sends real Gaussian states into real Gaussian states.
Nevertheless,  as the operator sum representation of a  Gaussian channel is not easily obtained and Gaussian states do not form a convex set,  the analogues of conditions (I3) and (I4) for a Gaussian imaginarity measure may not be suitable. Thus, as stated in \cite{Xu1}, the fundamental requirements for a Gaussian imaginarity measure ${\mathcal I}^G$ should be

(GI1) Faithfulness: ${\mathcal I}^G(\rho)\geq 0$ for all Gaussian states, and ${\mathcal I}^G(\rho)= 0$ iff $\rho$ is a real Gaussian state.

(GI2) Nonincreasing under real Gaussian operations: ${\mathcal I}^G(\Phi(\rho))\leq {\mathcal I}^G(\rho)$ for any real Gaussian channel $\Phi$ and any Gaussian state $\rho$.

Xu  in \cite{Xu1} and \cite{Xu2} respectively proposed two  Gaussian imaginarity measures $M_F$ and $M_{T,\mu}$, based on fidelity and Tsallis relative entropy for any $n$-mode Gaussian states, and showed that they satisfy the properties (GI1) and (GI2).
Thus,  with real Gaussian states as free states and real Gaussian channels  as free operations,  the Gaussian imaginarity is also a quantum resource.

However, the two Gaussian imaginarity measures $M_F$ and $M_{T,\mu}$ mentioned above are complicated  to compute due to their intricate formulas. As a result, when it comes to applying Gaussian imaginarity in the context of quantum information, a crucial task is to construct   quantifications of  multi-mode Gaussian  imaginarity that are easy to calculate. Note that a Gaussian state  is completely and conventionally described by its displacement vector (mean) and covariance matrix. The main purpose of this paper is to propose a Gaussian imaginarity measure $\mathcal{I}^{G_n}$  for any $n$-mode CV systems,  which is determined solely and formulated simply by the displacement vector and the covariance matrix of a Gaussian state, and consequently,  can be easily calculated. This enable us  to study the dynamics behaviour of Gaussian imaginarity by utilizing $\mathcal{I}^{G_n}$. Note that, the conception of Gaussian imaginarity is in term of single partite system, which is not a Gaussian quantum correlation. However, we show further that the Gaussian imaginarity measure $\mathcal{I}^{G_n}$ satisfies all requirements as a measure of multipartite multi-mode Gaussian correlation \cite{HLQ22}. This reveals that $n$-mode Gaussian imaginarity is also a multipartite multi-mode Gaussian correlation, and, together with the measure $\mathcal{I}^{G_n}$, is a multipartite Gaussian quantum resource.

The article is organized as follows. In Section II, we present  preliminaries about Gaussian states, Gaussian quantum channels and Gaussian imaginarity. Section III is devoted to introducing a quantification $\mathcal{I}^{G_n}$ for the imaginarity of Gaussian states, which is defined by the displacement vectors and covariance matrices of these states. We show that this quantification satisfies faithfulness and monotonicity under real Gaussian channels, and thus is a Gaussian imaginarity measure. In Section IV, for single-mode and $n$-mode Gaussian states, we compare $\mathcal{I}^{G_n}$ with known Gaussian imaginarity measures $M_F$ and $M_{T,\mu}$. Our results indicate that our measure  is easier to calculate, especially for large $n$.   In Section V,  we study the behaviour of the Gaussian imaginarity in Gaussian noise environments by applying ${\mathcal I}^{G_n}$ for $(1+1)$-mode Gaussian states to the scenarios in Markovian environments. In Section VI, we prove that the Gaussian quantum imaginarity ${\mathcal I}^{G_n}$ is  also a  multipartite multi-mode Gaussian quantum correlation measure. Section VII is a short conclusion. Proofs of some  results are presented in Appendix.

\section{Preliminaries}

In this section, we recall some notions and notations of real Gaussian states, real Gaussian quantum channels and Gaussian imaginarity measures.

\subsection{Real Gaussian states}

Consider an $n$-mode CV system with state space $H=H_1\otimes H_2\otimes \cdots\otimes H_n$, where each $H_k \ (1\le k\le n)$ is an infinite-dimensional complex Hilbert space.
Denote the one-mode Fock basis by $\{|j_m\rangle\}_{j_m=0}^\infty\subset H_m$,  $m=1,2,\cdots,n$.  Then the $n$-mode Fock basis of $H$ is
$\{|j_1\rangle\otimes |j_2\rangle \otimes \cdots \otimes |j_n\rangle\}_{j_1,j_2,\cdots,j_n=0}^\infty$.

Denote by $\mathcal{S}(H)$ the set of all quantum states (that is, positive bounded linear operators with trace 1) on $H$. For any state $\rho\in\mathcal{S}(H)$,  its characteristic function $\chi_{\rho}$ is defined as
$$\chi_\rho(z)={\rm Tr}(\rho W(z)),$$
where  $z=(x_{1}, y_{1}, \cdots, x_{n}, y_{n})^{\rm T}\in{\mathbb R}^{2n}$,
$W(z)=\exp(i{R^{\rm T}}z)$ is the Weyl displacement operator and
$R
=({R}_1,{R}_2,\ldots,{R}_{2n})=(\hat{Q}_1,\hat{P}_1,\ldots, \hat{Q}_n,\hat{P}_n)$.
Here, as usual, $\hat{Q}_k=\hat{a}_k+\hat{a}^\dagger_k$ and $\hat{P}_k=-i(\hat{a}_k-\hat a^\dag_k)$ $(k=1,2,\ldots,n)$ respectively stand for the position and momentum operators, with
 $\hat{a}^\dagger_k$ and $\hat{a}_k$  the creation and annihilation operators in the $k$-th mode $H_k$ satisfying the Canonical Commutation Relation (CCR)
$$[\hat{a}_k,\hat{a}_l^\dag]=\delta_{kl}I\ {\rm and}
\ [\hat{a}_k^\dag,\hat{a}_l^\dag]=[\hat{a}_k,\hat{a}_l]=0,\ \
k,l=1,2,\ldots,n.$$
Assume that $\rho$ has finite second moment. The displacement vector (or mean) ${\bar d}={\bar d}_\rho=(d_1, d_2, \ldots, d_{2n})^{\rm T}$ of $\rho$ is given by
$$\begin{array}{rl}{\bar d}=&(\langle\hat R_1 \rangle, \langle\hat R_2\rangle, \ldots ,\langle\hat R_{2n} \rangle)^{\rm T}\\
=&({\rm Tr}(\rho\hat R_1), {\rm Tr}(\rho \hat R_2), \ldots, {\rm Tr}(\rho \hat
R_{2n}))^{\rm T}\in{\mathbb R}^{2n},
\end{array}$$
and the covariance matrix (CM) $\nu=\nu_\rho=(\nu_{kl})\in {\mathcal M}_{2n}(\mathbb R)$ of $\rho$ is defined as
$$\nu_{kl}=\frac{1}{2}{\rm Tr}[\rho(\Delta\hat{R}_k\Delta\hat{R}_l+\Delta\hat{R}_l\Delta\hat{R}_k)],$$ where $\Delta\hat{R}_k=\hat{R}_k-\langle\hat{R}_k\rangle$ \cite{BV}.
It should be noted that a CM $\nu$ is real symmetric and must satisfy the uncertainty principle
$$\nu+i\Delta_n\ge0,$$ where $\Delta_n=\oplus_{i=1}^n\Delta_i$ with $\Delta_i=\left(
\begin{array}{cc}
0 & 1 \\
-1 & 0\\
\end{array}\right)$ for each $i$ \cite{SMB}. In addition, $\nu\ge 0$ as $\nu+i\Delta_n\ge0$.

Recall that $\rho$ is called a Gaussian state  if
$$\chi_\rho(z)={\rm Tr}(\rho W(z))=\exp(-\frac{1}{2}z^{\rm T}\nu z+i{\bar
	d}^{\rm T}z);$$
and is called a real Gaussian state if $\rho$ is Gaussian and satisfies
$$\langle k_1|\langle k_2|\cdots\langle k_n|\rho|l_1\rangle|l_2\rangle\cdots|l_n\rangle\in\mathbb{R}$$
for any Fock basis vectors $\{|k_1\rangle,|k_2\rangle,\ldots,|k_n\rangle$, $|l_1\rangle,|l_2\rangle,\cdots,|l_n\rangle\}$ \cite{Ser, Xu1}. It is shown in \cite[Theorem 1]{Xu1} that $\rho$ is real if and only if the displacement vector ${\bar d}$ and CM $\nu$ of $\rho$ fulfill the following conditions
\begin{equation}\label{eq1}
\begin{cases}d_{2k}=0\ \ {\rm for}\ \ k\in\{1,2,\ldots,n\},\\
	\nu_{2k-1,2l}=0\ \ {\rm for }\ \ k,l\in\{1,2,\ldots,n\}.\end{cases}
\end{equation}
Obviously, a Gaussian state $\rho$ is uniquely determined by  $\bar{d}$ and  $\nu$. So, we denote a Gaussian state $\rho$ by $\rho=\rho(\bar{d},\nu)$ sometimes.

\subsection{Real Gaussian quantum channels}

A quantum channel $\Phi$ on $\mathcal S(H)$ is   Gaussian if $\Phi$ sends Gaussian states into Gaussian states; and is real Gaussian if $\Phi$ is Gaussian and transforms real Gaussian states into real Gaussian states. Specifically, an $n$-mode Gaussian channel $\Phi$ can be represented as $\Phi(T,N,\bar{d}_0)$ in the following way \cite{CEG}: for arbitrary $n$-mode Gaussian state $\rho=\rho(\bar{d},\nu)\in{\mathcal S}(H)$, write $\Phi(\rho)=\rho(\bar{d}',\nu')$, then $$\bar{d}'=T\bar{d}+\bar{d}_0,\ \ \nu'=T\nu T^{\rm T}+N,$$ where, $\bar{d}_0=(d^0_1,d^0_2,\ldots,d^0_{2n})^{\rm T}\in{\mathbb R}^{2n}$, $T=(t_{ij})$ and $N=N^{\rm T}=(n_{ij})\ge0$ are $2n\times 2n$ real matrices satisfying
$$N+i\Delta_n-iT\Delta_n T^{\rm T}\ge 0.$$ It is shown  in \cite{Xu1} that a Gaussian channel $\Phi(T,N,\bar{d}_0)$ is real if and only if
\begin{equation}\label{eq2}
\begin{cases}d^0_{2k}=0\ \ {\rm for}\ \ k\in\{1,2,\ldots,n\},\\
n_{2k-1,2l}=0\ \ {\rm for }\ \ k,l\in\{1,2,\ldots,n\},\end{cases}\end{equation}
and either
\begin{equation}\label{eq3}
t_{2k,2l-1}=t_{2k,2l}=0 \ \ {\rm for} \ \ k,l\in\{1,2,\ldots,n\}
\end{equation}
or
\begin{equation}\label{eq4}
t_{2k-1,2l}=t_{2k,2l-1}=0 \ \ {\rm for} \ \ k,l\in\{1,2,\ldots,n\}.
\end{equation}
Furthermore, a real Gaussian channel $\Phi$ is called a completely real Gaussian channel if it satisfies Eqs.\eqref{eq2} and \eqref{eq3},
and is called a covariant real Gaussian channel if it fulfills  Eqs.\eqref{eq2} and \eqref{eq4}. Xu  proved that (ref. \cite[Theorem 4]{Xu1}), if $\Phi$ is a completely real Gaussian channel, then
$\Phi(\rho)$ is real for any Gaussian state $\rho$, that is, $\Phi$ is imaginarity breaking.

\subsection{Gaussian imaginarity measures}

In the framework of quantum resource theory, Xu  \cite{Xu1} proposed that a reasonable imaginarity measure $\mathcal{I}$ for $n$-mode Gaussian states should meet the   conditions (GI1) and (GI2) mentioned in the introduction section.

Assume that $\rho=\rho(\bar{d},\nu)$ is any $n$-mode Gaussian state with
$\rho^*=\rho(\bar{d^*},\nu^*)$ the conjugate of $\rho$.
In \cite{Xu1}, Xu gave a Gaussian imaginarity measure $M_F$  based on fidelity by
\begin{equation}\label{eq5}
M_F(\rho)=1-F(\rho,\rho^*),\end{equation}
where $F(\rho,\sigma)={\rm Tr}\sqrt{\sqrt{\rho}\sigma\sqrt{\rho}}$, and proved  that $M_F$ satisfies the conditions (GI1) and (GI2). In addition,  it is obtained in \cite{BBP} that

\begin{equation}\label{eq6}
M_F(\rho)=1-F_0(\rho,\rho^*)\exp(-\frac{1}{4}(\bar d-\bar {d^*})^{\rm T}(\nu+\nu^*)^{-1}(\bar d-\bar {d^*}),
\end{equation}
where
$$\begin{cases}
F_0(\rho,\rho^*)=\dfrac{F_{{\rm tot}}}{\sqrt[4]{\det(\nu+\nu^*)}},\\
F_{{\rm tot}}^4=\det\left((\sqrt{1-W_{{\rm aux}}^{-2}+I_{2n}})W_{{\rm aux}}i\Delta_n\right),\\
W_{{\rm aux}}=-(W_1+W_2)^{-1}(I_{2n}+W_2W_1),\\
W_1=-2\nu i\Delta_n, \ \ W_2=-2\nu^* i\Delta_n,
\end{cases}$$
and $\det(\cdot)$ denotes the determinant.

Later, Xu in \cite{Xu2} proposed another  Gaussian imaginarity measure $M_{T,\mu}$ based on Tsallis relative entropy by
\begin{equation}\label{eq7}
M_{T,\mu}(\rho)=1-{\rm Tr}[\rho^\mu(\rho^*)^{1-\mu}],
\end{equation}
where $\mu\in(0,1)$. It was shown \cite{Xu2} that $M_{T,\mu}$ also fulfills (GI1) and (GI2). For the CM $\nu$ of $\rho$, there exists some $2n\times2n$ real and symplectic
matrix $S$ (i.e. $S$ satisfies $S\Delta_n S^{\rm T}=\Delta_n$) such that
$$\nu=S(\oplus^n_{l=1}\nu_lI_2)S^{\rm T}\ \ {\rm with}\ \ \nu_l\ge1.$$ Then the CM $\nu^*$ of $\rho^*$ satisfies
$$\nu^*=OS(\oplus^n_{l=1}\nu_lI_2)S^{\rm T}O\ \ {\rm with}\ \ O=\oplus^n_{l=1}\left(\begin{array}{cc}
	1 & 0 \\
	0 & -1 \\
\end{array}\right).$$
$M_{T,\mu}(\rho)$ can be represented  as
\begin{equation}\label{eq8}
\begin{aligned}
M_{T,\mu}(\rho)&=1-\frac{2^n\prod_{l=1}^n\frac{(1-e^{-\eta_l})}{(1-e^{-\mu\eta_l})(1-e^{-(1- \mu)\eta_l})}}{\sqrt{\det(\nu^{(\mu)}+\nu^{*(1-\mu)})}}\\
&\times\exp\left\{-\frac{1}{2}(\bar{d}-\bar{d}^*)^{\rm T}(\nu^{(\mu)}+\nu^{*(1-\mu)})^{-1}(\bar{d}-\bar{d}^*)\right\},
\end{aligned}\end{equation}
where  $\eta_l=\ln\frac{\nu_l+1}{\nu_l-1}\ge0$ for $l\in\{1,2,\cdots,n\}$,  $\nu^{(\mu)}=S(\oplus^n_{l=1}\nu_l^{(\mu)}I_2)S^{\rm T}$ with
$\nu^{(\mu)}_l=\frac{2}{1-(\frac{\nu_l-1}{\nu_l+1})^{\mu}}-1$ for each $l$, and $\nu^{*(1-\mu)}=O\nu^{(1-\mu)}O$.

\section{A Gaussian imaginarity measure based on displacement vector and CM}

In this section, we  propose a more easily computable Gaussian imaginary measure for any $n$-mode CV systems based on the displacement vectors and CMs of Gaussian states.

\subsection{The single-mode systems}

To make the idea more clear we first  consider the single-mode CV system with the state space $H=H_1$. In this case, for any Gaussian state $\rho\in\mathcal{S}(H)$, the displacement vector and CM of $\rho$ are given by
\begin{equation*}\label{eq9}
\bar{d}=(d_1,d_2)^{\rm T}\ \ {\rm and}\ \ \nu=\left(\begin{array}{cc}
\nu_{11} & \nu_{12} \\
\nu_{12} & \nu_{22} \\
\end{array}
\right).\end{equation*}

\begin{definition}\label{Df1}
 For any one-mode Gaussian state $\rho=\rho(\bar{d},\nu)\in{\mathcal S}(H)$,  define
$$
\mathcal{I}^{G}(\rho)=1-\frac{\det(\nu)}{\det(Q_1\nu Q_1^{\rm T})\det(Q'_1\nu Q_1'^{\rm T})}+h(\|Q_1'\bar{d}\|_1),
$$
where $Q_1^{\rm T}=(1, 0)^{\rm T}\in\mathbb R^2$, $Q_1'^{\rm T}=(0,1)^{\rm T}\in\mathbb R^2$ and $h:[0,+\infty)\rightarrow \{0,1\}$ is a function with $h(t)=0$ if $t=0$ and $h(t)=1$ if $t\ne0$.
\end{definition}

It is clear that $Q_1'\bar{d}=d_2$, $Q_1\nu Q_1^{\rm T}=\nu_{11}$, $Q'_1\nu Q_1'^{\rm T}=\nu_{22}$ and $0<\det(\nu)\le\nu_{11}\nu_{22}$. So  $\mathcal{I}^{G}(\rho)\ge 0$ and
\begin{equation}\label{eq10}
\mathcal{I}^{G}(\rho)=\left\{\begin{array}{lr} 1-\frac{\nu_{11}\nu_{22}-\nu^2_{12}}{\nu_{11}\nu_{22}}=\frac{v_{12}^2}{v_{11}v_{22}}& {\rm if}\ d_2=0,\\
2-\frac{\nu_{11}\nu_{22}-\nu^2_{12}}{\nu_{11}\nu_{22}}=1+\frac{v_{12}^2}{v_{11}v_{22}} & {\rm if}\ d_2\ne0. \end{array}\right. \end{equation}
Notice that, $0\le\mathcal{I}^{G}<1$ if $d_2=0$ and $1\le\mathcal{I}^{G} \leq2$ if $d_2\not=0$. Obviously, the bounds 1 and 2 are tight.

By Eq.\eqref{eq1},  we see that $\rho$ is real if and only if $d_2=\nu_{12}=0$. It follows from Eq.\eqref{eq10} that $\mathcal{I}^{G}(\rho)=0$ if $\rho$ is real. Conversely, if $\mathcal{I}^{G}(\rho)=0$, then $1+h(|d_2|)=\frac{\nu_{11}\nu_{22}-\nu^2_{12}}{\nu_{11}\nu_{22}}\le1$, which implies that $h(|d_2|)=0$ and $\frac{\nu_{11}\nu_{22}-\nu^2_{12}}{\nu_{11}\nu_{22}}=1$. So  $d_2=\nu_{12}=0$, that is, $\rho$ is real. Thus, we have proved  the following result.

\begin{theorem}\label{thm2}
 For any single-mode Gaussian state $\rho\in{\mathcal S}(H)$, $\mathcal{I}^{G}(\rho)\geq 0$ and  $\mathcal{I}^{G}(\rho)=0$ if and only if $\rho$ is real.
\end{theorem}

Next, take any  real one-mode Gaussian channel $\Phi=\Phi(T,N,\bar{d}_0)$.  For any Gaussian state $\rho=\rho(\bar{d},\nu)$, by Eqs.\eqref{eq2}-\eqref{eq4}, one has $\Phi(\rho)=\rho(\bar{d}', \nu')$ with
$$\bar{d}'=T\bar{d}+\bar{d}_0\ \ {\rm and}\ \  \nu'=T\nu T^{\rm T}+N,$$  where $\bar{d}_0=(d_1^0,0)^{\rm T}\in\mathbb{R}^2$, $N=\left(\begin{array}{cc}
	n_{11} & 0 \\ 0  & n_{22} \\
\end{array}\right)\geq 0$, and either
$T=\left(\begin{array}{cc}
	t_{11} & t_{12} \\ 0  & 0 \\
\end{array}\right)$ or $T=\left(\begin{array}{cc}
	t_{11} & 0 \\ 0  & t_{22} \\
\end{array}\right)$. If $T$ has the first form, then $\Phi$ is a completely real Gaussian channel. So $\Phi(\rho)$ is real, and  $\mathcal{I}^{G}(\Phi(\rho))=0\le\mathcal{I}^{G}(\rho)$. If $T$ has the second form, then
$$\bar{d}'=(t_{11}d_1+d^0_1,t_{22}d_2)^{\rm T}$$and$$ \nu'=\left(\begin{array}{cc}
	t^2_{11}\nu_{11}+n_{11} & t_{11}t_{22}\nu_{12}\\ t_{11}t_{22}\nu_{12} & t^2_{22}\nu_{22}+n_{22} \\
\end{array}\right).$$
Hence
$$\begin{array}{rl}&\mathcal{I}^{G}(\Phi(\rho))\\
	= & 1-\dfrac{\det(\nu')}{\det(Q_1\nu'Q_1^{\rm T})\det(Q'_1\nu'Q_1'^{\rm T})}+h(\|Q_1'\bar{d}'\|_1)\\
	= & 1-\dfrac{(t^2_{11}\nu_{11}+n_{11})(t^2_{22}\nu_{22}+n_{22})-(t_{11}t_{22}\nu_{12})^2}
	{(t^2_{11}\nu_{11}+n_{11})(t^2_{22}\nu_{22}+n_{22})}\\&+h(|t_{22}d_2|)\\
	= & \dfrac{(t_{11}t_{22}\nu_{12})^2}
	{(t^2_{11}\nu_{11}+n_{11})(t^2_{22}\nu_{22}+n_{22})}+h(|t_{22}d_2|)\\
	\le & \dfrac{(t^2_{11}\nu_{11}+n_{11})(t^2_{22}\nu_{22}+n_{22})\nu^2_{12}}
	{(t^2_{11}\nu_{11}+n_{11})(t^2_{22}\nu_{22}+n_{22})\nu_{11}\nu_{22}}+h(|d_2|)\\
	= & \dfrac{\nu^2_{12}}{\nu_{11}\nu_{22}}+h(|d_2|)=\mathcal{I}^{G}(\rho).
\end{array}
$$

This leads to the following conclusion.

\begin{theorem}\label{thm3}  For any single-mode Gaussian state $\rho\in{\mathcal S}(H)$ and any real Gaussian channel $\Phi$, we have
$$\mathcal{I}^{G}(\Phi(\rho))\le \mathcal{I}^{G}(\rho).$$
\end{theorem}

The Theorems 2 and 3 ensure that $\mathcal{I}^{G}$ satisfies the requirements of faithfulness (GI1) and monotonicity (GI2), and therefore,   serves as a suitable Gaussian
imaginary measure for single-mode CV system.

\subsection{ The multi-mode systems}

In this subsection, we generalize Definition \ref{Df1} to the scenario of  $n$-mode CV systems for any $n\geq 1$.
Assume that $H=H_1\otimes H_2\otimes \cdots\otimes H_n$ and $h$ is the function defined in  Definition \ref{Df1}.

\begin{definition}\label{Df4}
 For any $n$-mode Gaussian state $\rho=\rho(\bar{d},\nu)\in\mathcal{S}(H)$ with  $\bar{d}=(d_1,d_2,\ldots,d_{2n})^{\rm T}$ and $\nu=(\nu_{kl})_{2n\times 2n}$, we define $\mathcal{I}^{G_n}(\rho)$ as
\begin{widetext}
\begin{equation}\label{eq11}
\mathcal{I}^{G_n}(\rho)=1-\frac{\det(\nu)}{\det(Q_nP_n\nu P_n^{\rm T}Q_n^{\rm T})\det(Q'_nP_n\nu P_n^{\rm T}Q_n'^{\rm T})}+h(\|Q_n'P_n\bar{d}\|_1),
\end{equation}
\end{widetext}
where $P_n=(p_{kl})_{2n\times 2n}\in\mathcal{M}_{2n}(\mathbb{R})$ is  a permutation matrix satisfying $p_{k,2k-1}=p_{n+k,2k}=1$ for $k\in\{1,2,\ldots,n\}$
and other elements  0, $Q_n=(I_n,0)_{n\times2n}$ and $Q'_n=(0,I_n)_{n\times2n}$.
\end{definition}

By the definition of $P_n$ in Definition \ref{Df4}, one gets
\begin{widetext}
	$$P_n\nu P_n^{\rm T}=\left(
                    \begin{array}{cccc|cccc}
                      \nu_{11}     & \nu_{13}     & \cdots  & \nu_{1,2n-1}     & \nu_{12}      & \nu_{14}      & \cdots  & \nu_{1,2n} \\
                      \nu_{13}     & \nu_{33}     & \cdots  & \nu_{3,2n-1}     & \nu_{23}      & \nu_{34}      & \cdots  & \nu_{3,2n} \\
                      \vdots       & \vdots       &  \ddots       & \vdots           & \vdots        & \vdots        &  \ddots       & \vdots \\
                      \nu_{1,2n-1} & \nu_{3,2n-1} & \cdots  & \nu_{2n-1,2n-1}  & \nu_{2n-1,2}  & \nu_{2n-1,4}  & \cdots  & \nu_{2n-1,2n} \\
                      \hline
                      \nu_{12}     & \nu_{23}     & \cdots  & \nu_{2n-1,2}     & \nu_{22}      & \nu_{24}      & \cdots  & \nu_{2,2n} \\
                      \nu_{14}     & \nu_{34}     & \cdots  & \nu_{2n-1,4}     & \nu_{24}      & \nu_{44}      & \cdots  & \nu_{4,2n} \\
                      \vdots       & \vdots       &   \ddots      & \vdots           & \vdots        & \vdots        &  \ddots       & \vdots \\
                      \nu_{1,2n}   & \nu_{3,2n}   & \cdots  & \nu_{2n-1,2n}    & \nu_{2,2n}    & \nu_{4,2n}    & \cdots  & \nu_{2n,2n} \\
                    \end{array}
                  \right)= \left(
                    \begin{array}{cc}
                      A_{11}     & A_{12} \\
                      A^{\rm T}_{12} & A_{22} \\
                    \end{array}
                  \right),$$
                  \end{widetext}
where $A_{kl}\in\mathcal{M}_{n}(\mathbb{R})$, $k,l\in\{1,2\}$. Furthermore, a straightforward calculation gives
 $$\begin{cases}Q_nP_n\nu P_n^{\rm T}Q_n^{\rm T}=A_{11},\ \ Q'_nP_n\nu P_n^{\rm T}Q_n'^{\rm T}=A_{22},\\
 P_n\bar{d}=(d_1,d_3,\ldots,d_{2n-1},d_2,d_4,\ldots,d_{2n})^{\rm T},\\
Q_nP_n\bar{d}=(d_1,d_3,\ldots,d_{2n-1})^{\rm T}\in\mathbb{R}^n,\\
Q_n'P_n\bar{d}=(d_2,d_4,\ldots,d_{2n})^{\rm T}\in\mathbb{R}^n.
\end{cases}$$ Consequently, Eq.\eqref{eq11} can be expressed as
\begin{equation}\label{eq12}
\begin{array}{rl}
\mathcal{I}^{G_n}(\rho)
=&1-\frac{\det(\nu)}{\det(A_{11})\det(A_{22})}\\&+h(|d_2|+|d_4|+\cdots+|d_{2n}|).\end{array}
\end{equation}

It is obvious that, for $n=1$, ${\mathcal I}^{G_1}$ is the same as ${\mathcal I}^G$ defined in preceding subsection.

The following Theorems \ref{thm5} and \ref{thm6} rigorously establish that $\mathcal{I}^{G_n}$ is a well-defined measure of multi-mode Gaussian imaginarity, fulfilling both the faithfulness (GI1) and monotonicity (GI2).

\begin{theorem}\label{thm5}
For any $n$-mode Gaussian state $\rho\in\mathcal{S}(H)$, $\mathcal{I}^{G_n}(\rho)\geq 0$ and $\mathcal{I}^{G_n}(\rho)=0$ if and only if $\rho$ is real.
\end{theorem}

\begin{theorem}\label{thm6}
 For any $n$-mode Gaussian state $\rho\in\mathcal{S}(H)$ and any $n$-mode real Gaussian channel $\Phi$, we have $\mathcal{I}^{G_n}(\Phi(\rho))\le\mathcal{I}^{G_n}(\rho)$.
\end{theorem}

We present the proofs of Theorems \ref{thm5} and \ref{thm6} in Appendix.

\section{Comparing  $\mathcal{I}^{G_n}$  with $M_F$ and $M_{T,\mu}$}

In this section, we conduct comparative analyses of $\mathcal{I}^{G_n}$ with two established Gaussian imaginarity measures: the fidelity-based Gaussian imaginarity measure $M_F$ defined  in \cite{Xu1} (see Eq.\eqref{eq5}), and the Tsallis-type relative entropy imaginarity measure $M_{T,\mu}$ specified in \cite{Xu2} (see Eq.\eqref{eq7}).

Consider any one-mode Gaussian state $\rho$ characterized by the displacement vector $\bar{d}=(d_1,d_2)^{\rm T}$ and the CM $\nu=\left(\begin{array}{cc}
\nu_{11} & \nu_{12} \\
\nu_{12} & \nu_{22} \\
\end{array}
\right)$.
Note that $\rho(\nu,\bar{d})$ has the thermal decomposition
\begin{equation}\label{eqaa}
\rho(\nu,\bar{d})=D(\alpha)S(\zeta)\rho_\mathrm{th}S(-\zeta)D(-\alpha),
\end{equation}
where $D(\alpha)=\exp(\alpha\hat{a}^{\dagger}-\alpha^*\hat{a})$ is the one-mode displacement operator, $\alpha, \zeta$ are complex numbers, $\zeta=|\zeta|e^{i\theta}$ is the polar form,  $S(\zeta)=\exp\left[\frac{1}{2}(\zeta^*\hat{a}^2-\zeta\hat{a}^{\dagger 2})\right]$ is the one-mode squeezing operator, $\rho_\mathrm{th}$ is the single-mode thermal state \cite{Oli,MT,Adm}, that is
$$\rho_\mathrm{th}=\frac{1}{n_\mathrm{th}+1}\sum^{\infty}_{n=0}(\frac{{n_\mathrm{th}}}{{n_\mathrm{th}}+1})^n|n\rangle\langle n|.$$
Then the CM and the displacement vector of $\rho(\nu,\bar{d})$ can be written respectively as
\begin{widetext}
$$\nu=(1+2{n_\mathrm{th}})\begin{pmatrix}
\cosh(2|\zeta|) + \cos\theta\sinh(2|\zeta|) & \sin\theta\sinh(2|\zeta|) \\
\sin\theta\sinh(2|\zeta|) & \cosh(2|\zeta|) - \cos\theta\sinh(2|\zeta|)
\end{pmatrix}$$
\end{widetext}
and $\bar{d}=(2\mathrm{Re}\alpha, 2\mathrm{Im}\alpha)^{\rm T}$. Thus,  Eq.\eqref{eq5},  Eq.\eqref{eq7} and Eq.\eqref{eq10} can be respectively reduced to
\begin{widetext}

\begin{equation}\label{eq13}
M_F(\rho) = 1 - \dfrac{
    \exp\left(
        -\dfrac{2\, (\mathrm{Im}\alpha)^2}{(2n_{\mathrm{th}}+1)\bigl( \cosh(2|\zeta|)-\cos\theta \sinh(2|\zeta|) \bigr)}
    \right)}{\sqrt{ \sqrt{
(2n_{\mathrm{th}} + 1)^2 \bigl( 1 + \sin^2\theta \sinh^2(2|\zeta|) \bigr) + 4n_{\mathrm{th}}^2 (n_{\mathrm{th}} + 1)^2}-2n_{\mathrm{th}}(n_{\mathrm{th}} + 1)}}
\end{equation}

\begin{equation}\label{eq14}
\begin{aligned}
M_{T,\mu}(\rho) &= 1 - \frac{2(1-x)}{(1-x^{\mu})(1-x^{(1-\mu)})\sqrt{\left(\frac{2(1-x)}{(1-x^{\mu})(1-x^{(1-\mu)})}\right)^2
+4\left(\frac{(1+x^{\mu})(1+x^{(1-\mu)})}{(1-x^{\mu})(1-x^{(1-\mu)})}\right)\sin^2\theta \sinh^2(2|\zeta|)}} \\
&\quad \times \exp\left\{-\frac{4(\mathrm{Im}\,\alpha)^2(1-x)(\cosh(2|\zeta|)+\cos\theta\sinh(2|\zeta|))}
{\frac{(1-x)^2}{(1-x^{\mu})(1-x^{(1-\mu)})}
+(1+x^{\mu})(1+x^{(1-\mu)})\sin^2\theta \sinh^2(2|\zeta|)}\right\},
\end{aligned}
\end{equation}
with $x=\frac{n_\mathrm{th}}{n_\mathrm{th}+1}$, and
\begin{equation}\label{eqbb}
	\mathcal{I}^{G}(\rho)=
	\left\{\begin{array}{rr} 1-\frac{1}{1+\sin^2\theta\sinh^2(2|\zeta|)} & {\rm if}\ \alpha\in\mathbb{R},\\
		2-\frac{1}{1+\sin^2\theta\sinh^2(2|\zeta|)} & {\rm if}\ \alpha\notin\mathbb{R}.
		\end{array}\right.
	\end{equation}

\end{widetext}

Obviously, $\mathcal I^G(\rho)$ has the simplest expression and is independent of the average photon number $n_{\rm th}$. This observation also reveals that the single-mode  Gaussian imaginarity is  independent of the average photon number  of Gaussian states, which is a new discovery. It is worth noting that the difficulty in $M_F$ lies in computing the inverse of the matrix, while the challenge in $M_{T,\mu}$ involves calculating the symplectic eigenvalues of the matrix, as well as its inverse and square root. As $n$ grows larger, computing its inverse and square root becomes increasingly difficult.

\begin{example} Let $\rho=|\alpha\rangle\langle\alpha|$ be the single-mode Glauber coherent state, that is, $n_\mathrm{th}=0$ and $\zeta=0$ in Eq.\eqref{eqaa}, where
$$|\alpha\rangle=D(\alpha)|0\rangle=e^{-\frac{|\alpha|^2}{2}}\sum^{\infty}_{n=0}\frac{\alpha^n}{\sqrt{n!}}|n\rangle$$with $ \alpha \ {\rm any \ complex \ number}.$
\end{example}

\begin{figure}[hbp]
\centering
\includegraphics[width=1\linewidth,keepaspectratio]{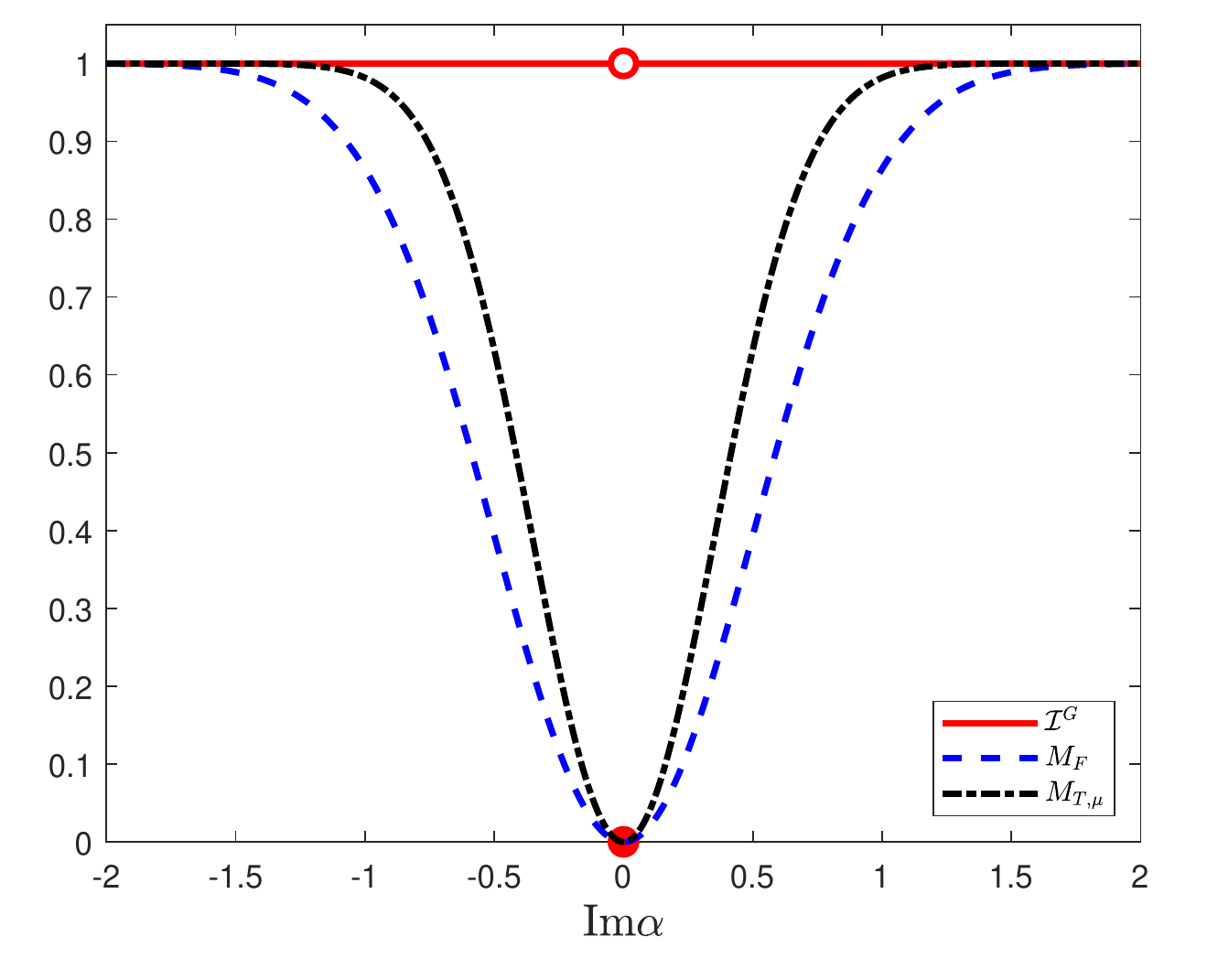}
\caption{\footnotesize $\mathcal{I}^{G},M_F,M_{T,\mu}$ in Eqs.\eqref{eq16}-\eqref{eq15} as the functions of  ${\rm Im}\alpha$.}
\label{1}
\end{figure}

In this case,  $\bar{d}=(2{\rm Re}\alpha,2{\rm Im}\alpha)^{\rm T}$ and  $\nu=\left(\begin{array}{cc}
1 & 0 \\
0 & 1 \\
\end{array}
\right)$. By Eqs.\eqref{eq13}-\eqref{eqbb}, we can derive that

\begin{equation}\label{eq16}
M_F(|\alpha\rangle\langle\alpha|)=1-e^{-2({\rm Im}\alpha)^2},
\end{equation}
\begin{equation}\label{eq17}
M_{T,\mu}(|\alpha\rangle\langle\alpha|)=1-e^{-4({\rm Im}\alpha)^2}
\end{equation}
and
\begin{equation}\label{eq15}
	\mathcal{I}^{G}(|\alpha\rangle\langle\alpha|)=h(2|{\rm Im}\alpha|)=\left\{\begin{array}{rr} 0& {\rm if}\ \alpha\in\mathbb{R},\\
		1 & {\rm if}\ \alpha\notin\mathbb{R}. \end{array}\right.\end{equation}
Evidently, $\mathcal{I}^{G}(|\alpha\rangle\langle\alpha|)>M_{T,\mu}(|\alpha\rangle\langle\alpha|)>M_F(|\alpha\rangle\langle\alpha|)>0$ if $\alpha\notin\mathbb{R}$ and $\mathcal{I}^{G}(|\alpha\rangle\langle\alpha|)=M_{T,\mu}(|\alpha\rangle\langle\alpha|)=M_F(|\alpha\rangle\langle\alpha|)=0$ whenever $\alpha\in\mathbb{R}$. Figure \ref{1} gives the images of $\mathcal{I}^{G},M_F,M_{T,\mu}$ as functions of ${\rm Im}\alpha$.  We  see from Fig.\ref{1} that, if $|{\rm Im}\alpha|$ is too small, it is very difficult to identify the imaginarity of $|\alpha\rangle\langle\alpha|$ by utilizing  $M_F$ and $M_{T,\mu}$ as $M_F(\rho)$ and $M_{T,\mu}(\rho)$ are very close to zero; however, it is easy to adjudge that $|\alpha\rangle\langle\alpha|$   contains the imaginarity  by $\mathcal I^G$.   This reveals that $\mathcal{I}^{G}$ has more advantages than $M_F$ and $M_{T,\mu}$ in detecting whether a single-mode Gaussian state is real.

\begin{example}
Consider the single-mode squeezed state $|\zeta\rangle\langle\zeta|$ with
$$\begin{array}{rl}|\zeta\rangle=&\exp\big[\frac{1}{2}(\zeta^*\hat a^2-\zeta\hat a^{\dag 2})\big]|0\rangle\\=
&\frac{1}{\sqrt{\cosh|\zeta|}}\sum_{n=0}^{\infty}(e^{-i\theta}\tanh|\zeta|)^n\frac{\sqrt{(2n)!}}{2^nn!}
|2n\rangle,\end{array}$$
for every complex number $\zeta=|\zeta|e^{i\theta}$.
\end{example}

\begin{figure*}[tbp]
	\center
	\subfigure []
	{\includegraphics[width=8cm,height=6cm]{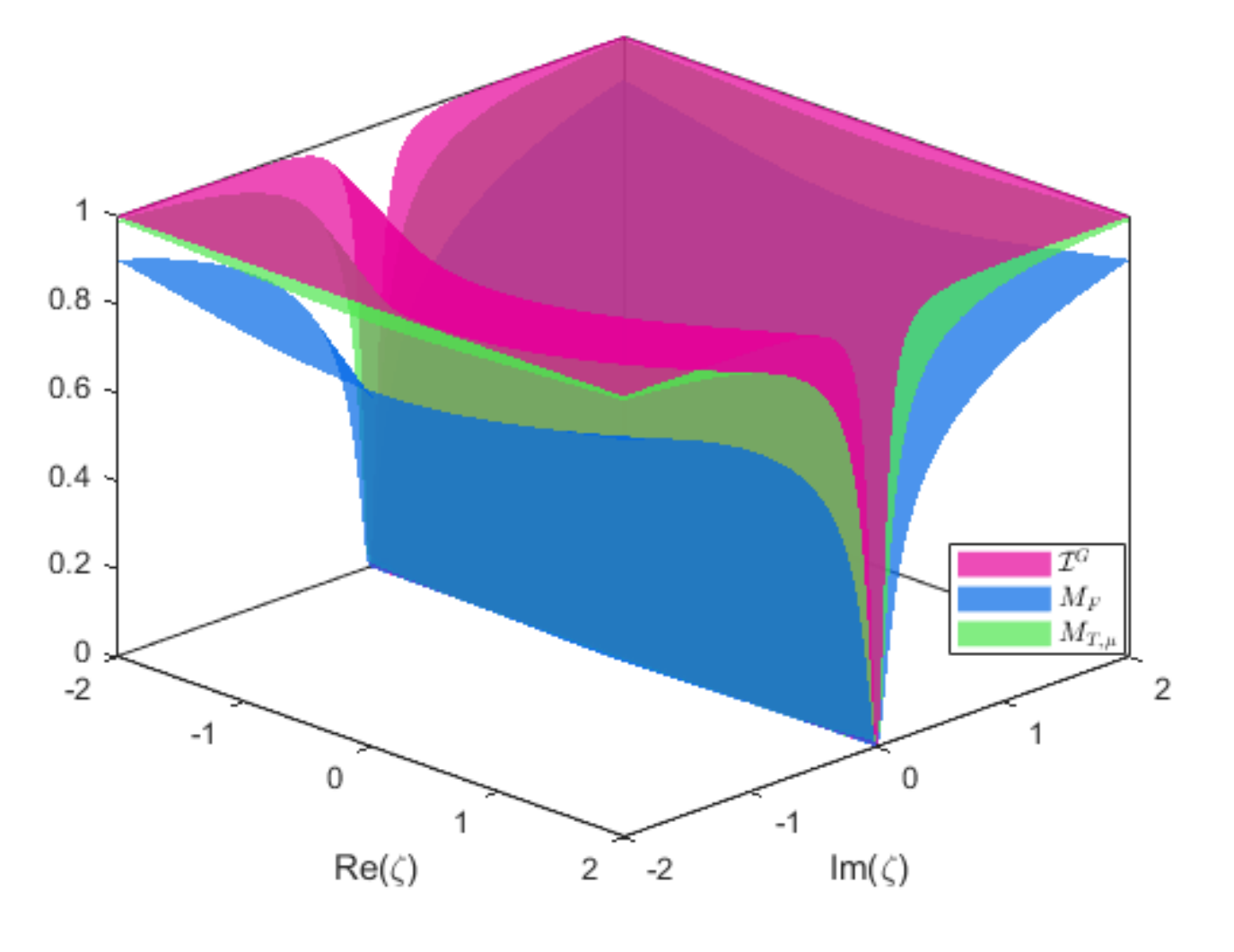}}
	\subfigure []
	{\includegraphics[width=8cm,height=6cm]{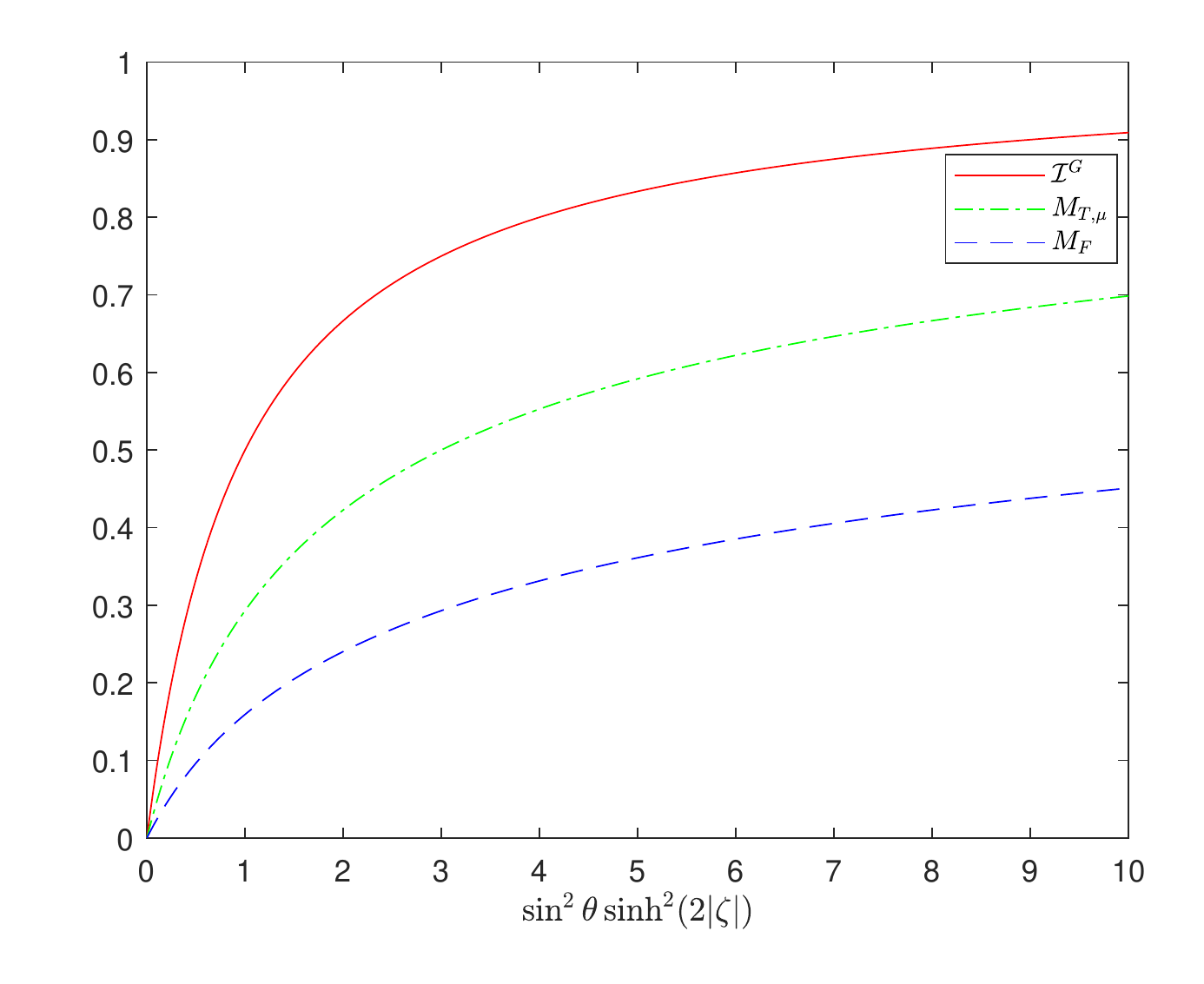}}
	\caption{\small Graphs of $\mathcal{I}^{G}, M_F, M_{T,\mu}$ in Eqs.\eqref{eq18}-\eqref{eq200}.  (a) Images of  $\mathcal{I}^{G}, M_F, M_{T,\mu}$ as functions of ${\rm Re}(\zeta)$ and ${\rm Im}(\zeta)$. (b) Images of  $\mathcal{I}^{G},M_F,M_{T,\mu}$ as functions of  $\sin^2\theta\sinh^2(2|\zeta|)$.}
\label{2}
\end{figure*}

In this case,  $n_\mathrm{th}=0$ and $\alpha=0$ in Eq.\eqref{eqaa}. The corresponding displacement vector has the form $\bar{d}_{|\zeta\rangle\langle\zeta|}=(0,0)^{\rm T}$,  while the CM is given by
$$\begin{array}{rl}
	&\nu_{|\zeta\rangle\langle\zeta|}\\
	=&{\footnotesize \left(\begin{array}{cc}
\cosh(2|\zeta|)+\cos\theta\sinh(2|\zeta|) & \sin\theta\sinh(2|\zeta|) \\
\sin\theta\sinh(2|\zeta|) & \cosh(2|\zeta|)-\cos\theta\sinh(2|\zeta|) \\
\end{array}
\right)}.\end{array}$$
Then Eqs.\eqref{eq13}-\eqref{eqbb} become to
\begin{equation}\label{eq18}
\mathcal{I}^{G}(|\zeta\rangle\langle\zeta|)=1-\frac{1}{1+\sin^2\theta\sinh^2(2|\zeta|)},
\end{equation}
\begin{equation}\label{eq19}
M_F(|\zeta\rangle\langle\zeta|)=1-\frac{1}{\sqrt[4]{1+\sin^2\theta\sinh^2(2|\zeta|)}}
\end{equation}
and
\begin{equation}\label{eq200}
M_{T,\mu}(|\zeta\rangle\langle\zeta|)=1-\frac{1}{\sqrt{1+\sin^2\theta\sinh^2(2|\zeta|)}}.
\end{equation}
It is apparent that $\mathcal{I}^{G}(|\zeta\rangle\langle\zeta|)> M_{T,\mu}(|\zeta\rangle\langle\zeta|)> M_F(|\zeta\rangle\langle\zeta|)>0$ if $\zeta\notin\mathbb{R}$ and $\mathcal{I}^{G}(|\zeta\rangle\langle\zeta|)=M_{T,\mu}(|\zeta\rangle\langle\zeta|)=M_F(|\zeta\rangle\langle\zeta|)=0$ whenever $\zeta\in\mathbb{R}$. We conduct a comprehensive comparison of Eqs.\eqref{eq18}-\eqref{eq200} using three-dimensional and two-dimensional graphs, see  Fig.\ref{2}.

Finally, we consider an $n$-mode Gassuain state.

\begin{example} Consider the $n$-mode Glauber coherent state
$$|\alpha\rangle=|\alpha_1\alpha_2\cdots\alpha_n\rangle=|\alpha_1\rangle\otimes|\alpha_2\rangle\otimes\cdots\otimes|\alpha_n\rangle,$$ where for each $l$ with $1\le l\le n$, $\alpha_l$ represents a single-mode Glauber coherent state. Specifically, it satisfies
$$\hat a_l|\alpha_l\rangle=\alpha_l|\alpha_l\rangle,\ \ 1\le l\le n,$$
or
$$|\alpha_l\rangle=D(\alpha_l)|0\rangle=e^{-\frac{|\alpha_l|^2}{2}}\sum^{\infty}_{n=0}\frac{\alpha_l^n}{\sqrt{n!}}|n\rangle, \ \ 1\le l\le n,$$
with $\alpha_l, \alpha_2,\ldots,\alpha_n$ being complex number.
\end{example}

Then $\bar{d}_{|\alpha\rangle\langle\alpha|}=(2{\rm Re}\alpha_1,2{\rm Im}\alpha_1,\ldots,2{\rm Re}\alpha_n,2{\rm Im}\alpha_n)^{\rm T}\in\mathbb{R}^{2n}$, and $\nu_{|\alpha\rangle\langle\alpha|}=I_{2n}$. By Eq.\eqref{eq12}, it can be readily verified that
$$\begin{array}{rl}
\mathcal{I}^{G_n}(|\alpha\rangle\langle\alpha|)&=h(2|{\rm Im}\alpha_1|+2|{\rm Im}\alpha_2|+\cdots+2|{\rm Im}\alpha_n|)\\
&=\left\{\begin{array}{rr} 0& {\rm if\ all}\ \alpha_i\in\mathbb{R},\\
1 & {\rm otherwise},\end{array}\right.
\end{array}$$
which implies that $|\alpha\rangle\langle\alpha|$ must contain quantum imaginarity when there exsits at least one $\alpha_i$ such that $\alpha_i\notin \mathbb R$.
Nevertheless, by Eq.\eqref{eq6} and Eq.\eqref{eq8}, computing the values of  $M_F(|\alpha\rangle\langle\alpha|)$ and $M_{T,\mu}(|\alpha\rangle\langle\alpha|)$  are complicated.

\section{Behavior of Gaussian imaginarity in Markovian environments}

As the multi-mode Gaussian imaginarity measure ${\mathcal I}^{G_n}$ is easily computed, it can be conveniently utilized to explore the behavior of Gaussian imaginarity within the context of system evolutionary processes. Here we illustrate this  by studying the behaviour of the Gaussian imaginarity in Gaussian noise environments by applying ${\mathcal I}^{G_n}$ to focus on the scenarios in Markovian environments for two-mode CV systems \cite{HPZ,ISM}.

The dynamics of any $(1+1)$-mode quantum state $\rho$ through a (Markovian) noisy
environment is given by the following master equation (\cite{Oli, GKS,LG,ISS,FOP})
$$\dot{\rho}=\sum_{k=1}^{2}\frac{\lambda}{2}\{(N+1)\mathcal{L}[\hat{a}_k]+N\mathcal{L}[\hat{a}_k^\dag]-M^{*}\mathcal{D}[\hat{a}_k]-M\mathcal{D}[\hat{a}_k^\dag]\}\rho,$$
where  $\hat{a}_k^\dag$ and $\hat{a}_k$ are the creation and annihilation
operators in the $k$-th mode, $\lambda$ is the overall damping rate, $N\in{\mathbb R}$ and $M\in{\mathbb C}$ represent the effective number of photons and the squeezing parameter of the bath, respectively, with $|M|^{2}\leq N(N+1)$, and $\mathcal{L}[A]\rho=2A\rho A^{^\dag}-A^{^\dag}A\rho-\rho A^{^\dag}A$ and $\mathcal{D}[A]\rho=2A\rho A-AA\rho-\rho AA$ are Lindblad superoperators. Time evolution imposed by the master equation preserves the Gaussian character of the state $\rho(t)$ and the
CM at time $t$ is given by (\cite{Oli, PIS,SPF})
\begin{equation}\label{eq20}
\nu_{\rho(t)}=e^{-\lambda t}\nu_{\rho(0)}+(1-e^{-\lambda t})\nu_{\rho(\infty)} \end{equation}
with
\begin{equation}\label{eq21}
\nu_{\rho(\infty)}=2\left(\begin{array}{cccc}\frac{1}{2}+L_{+}&{\rm Im} M &0&0\\ {\rm Im} M&\frac{1}{2}+L_{-}&0&0\\0&0&\frac{1}{2}+L_{+}&{\rm Im}M\\
0&0&{\rm Im}M&\frac{1}{2}+L_{-}\end{array}\right).
\end{equation}
Here, $N=n_\mathrm{th}({\rm cosh}^{2}(R)+{\rm sinh}^{2}(R))+{\rm sinh}^{2}R$, $M=-(2n_\mathrm{th}+1){\rm cosh}(R){\rm sinh}(R)e^{i\phi}$, $n_\mathrm{th}$ is the thermal photon number, $R$ is the squeezing parameter of the bath, $\phi$ is the squeezing phase and $L_{\pm}=N\pm {\rm Re}M$.

Also note that, by \cite{Oli}, the displacement vector $\bar{d}_{\rho(t)}$ of $\rho(t)$ is
\begin{equation}\label{eq22}
\bar{d}_{\rho(t)}=e^{-\frac{\lambda t}{2}}\bar{d}_{\rho(0)}.
\end{equation}

Firstly, assume that the initial Gaussian state $\rho(0)$ is a squeezed vacuum state with  the CM
$$\nu_{\rho(0)}=2\left(\begin{array}{cccc}{\rm cosh}2r&0&{\rm sinh}2r&0\\0&{\rm cosh}2r&0&-{\rm sinh}2r\\{\rm sinh}2r&0&{\rm cosh}2r&0\\
0&-{\rm sinh}2r&0&{\rm cosh}2r\end{array}\right),$$
where $r$ is the squeezing parameter of the state. As the displacement vector $\bar{d}_{\rho(0)}=0$, we see that $\rho(0)$ is a real stat by Eq.\eqref{eq1} and the displacement vector of $\rho(t)$ is $\bar{d}_{\rho(t)}=0$ for all $t$ by Eq.\eqref{eq22}.

We explore the behaviour of the Gaussian imaginarity contained in $\rho(t)$ by  $\mathcal I^{G_2}(\rho(t))$. Applying Eqs.\eqref{eq20}-\eqref{eq21}, one has
$$\nu_{\rho(t)}=\left(\begin{array}{cccc}
a_{+}&  c & b & 0\\
c &a_{-}& 0 & -b\\
b & 0 & a_{+} &c\\
0 & -b & c &a_{-}\end{array}\right),$$
where $a_{\pm}=2e^{-\lambda t}{\rm cosh}(2r)+(1-e^{-\lambda t})(1+2L_{\pm})$, $b=2e^{-\lambda t}{\rm sinh}(2r)$ and $c=2(1-e^{-\lambda t}){\rm Im}M$. From Eq.\eqref{eq12}, we get
\begin{equation*}\label{eq23}
\begin{aligned}
&\mathcal I^{G_2}(\rho(t))\\
=&1-\frac{\det(\nu_{\rho(t)})}{(a_{+}^2 - b^2)(a_{-}^2 - b^2)}\\
=&1-\frac{b^4 + c^4 + 2b^2c^2 + a_{+}^2a_{-}^2 - 2a_{+}c^2a_{-} - a_{+}^2b^2 - b^2a_{-}^2}{(a_{+}^2 - b^2)(a_{-}^2 - b^2)}.
\end{aligned}
\end{equation*}

\begin{figure*}[tbp]
\center
\subfigure []
{\includegraphics[width=8cm,height=6cm]{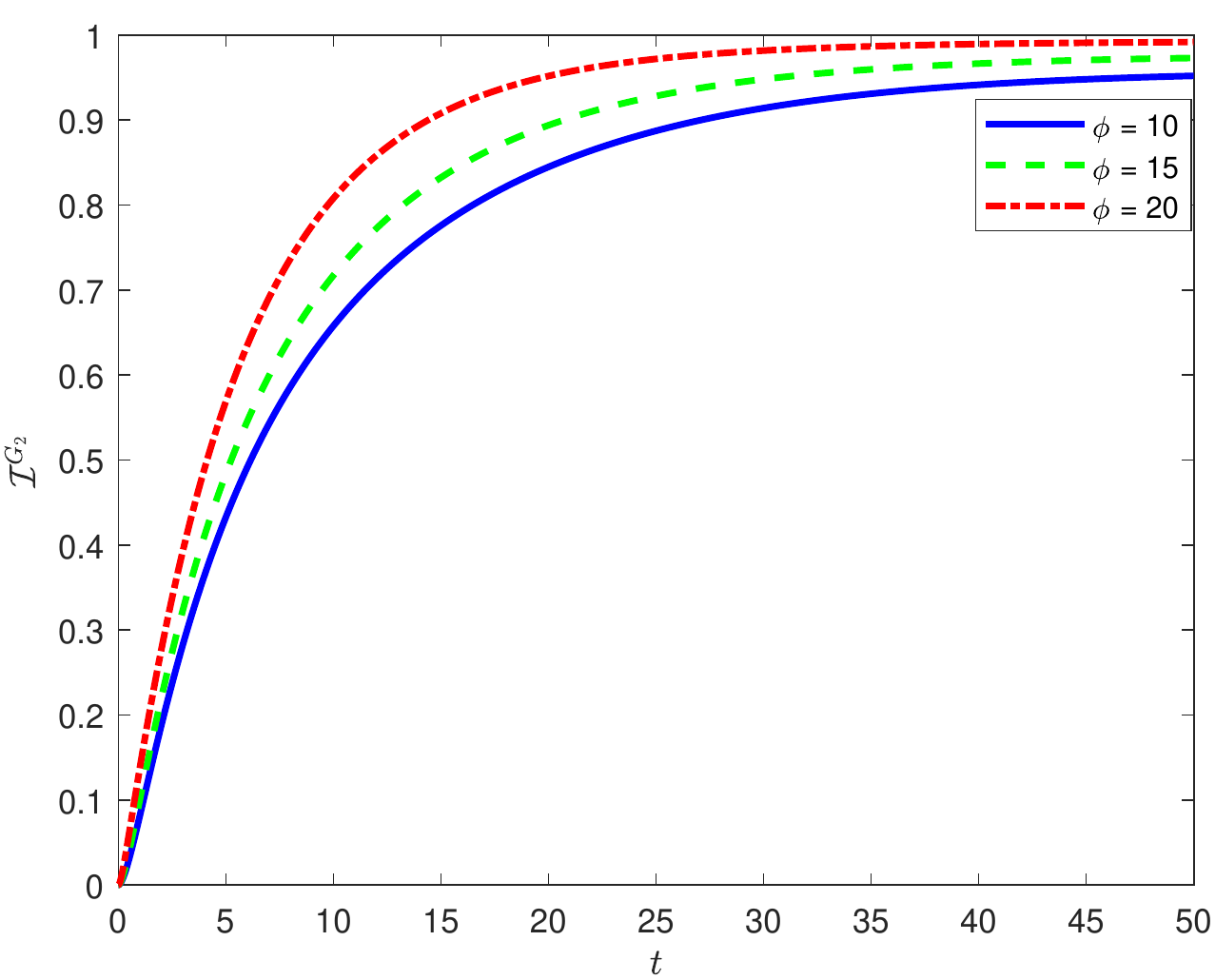}}
\subfigure []
{\includegraphics[width=8cm,height=6cm]{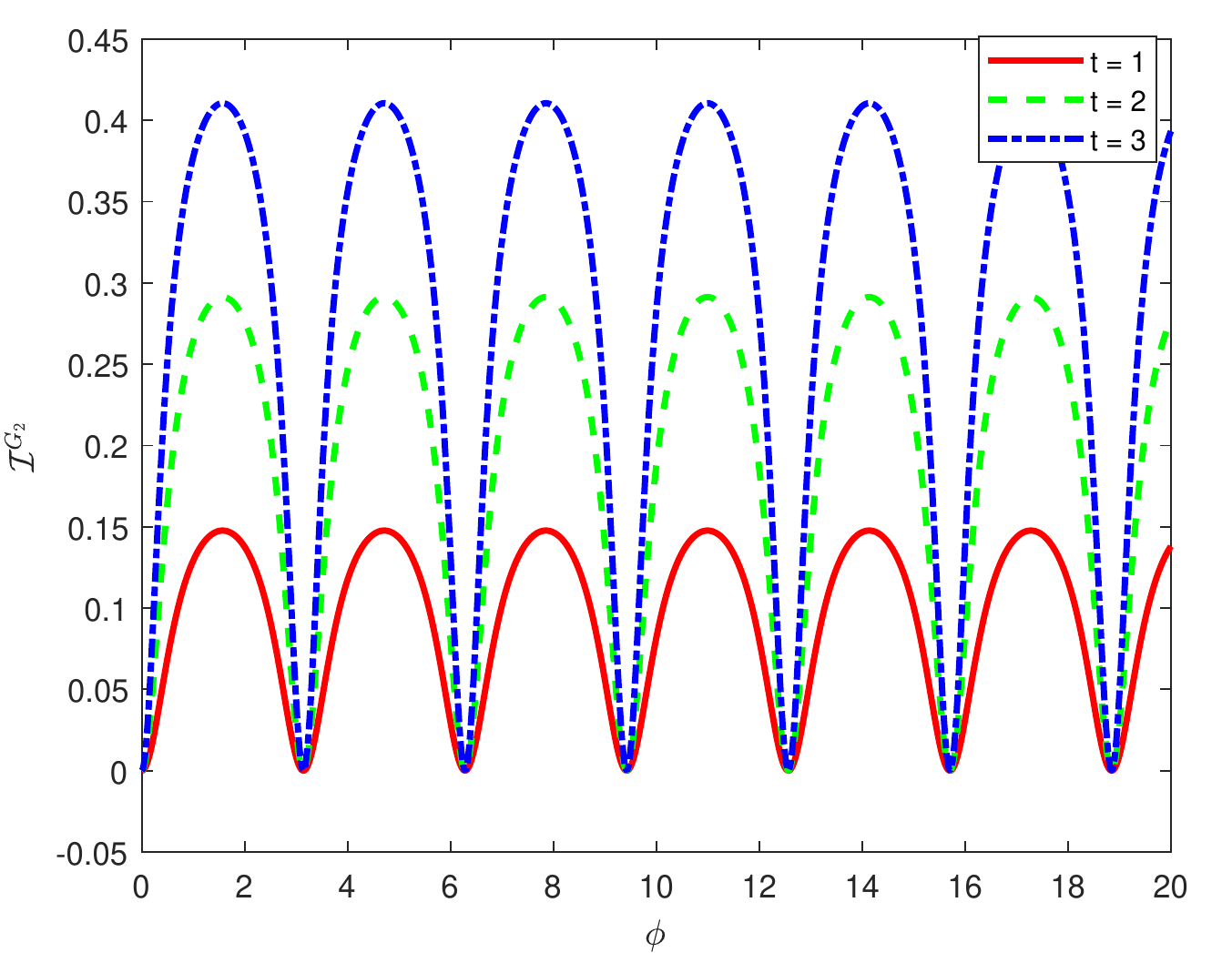}}
  \caption{\small Behavior of $\mathcal I^{G_2}(\rho(t))$, with $\rho(0)$ the squeezed vacuum state of $r=1$, as a function of the parameters $t$ and $\phi$ for fixed $n_\mathrm{th}=1.5, R = 1, \lambda=0.1$. (a) Plots of $\mathcal I^{G_2}$ as a function of the parameter $t$ for fixed $\phi= 10, 15,20$, respectively. (b) Plots of $\mathcal I^{G_2}$ as a function of the parameter $\phi$ for fixed $t=1,2,3$, respectively.}
\label{3}
\end{figure*}

To analysis the behavior, we fix the parameters $r$, $n_\mathrm{th}$, $R$, and $\lambda$ in $\nu_{\rho(t)}$, say $r=1$, $n_\mathrm{th}=1.5, R = 1, \lambda=0.1$ in Fig.\ref{3}, and then $\mathcal I^{G_2}(\rho(t))$ is a function of the time $t$ and the squeezing phase $\phi$. In Fig.\ref{3}(a),  we see that, for   $\phi=10,15,20$,  $\mathcal I^{G_2}(\rho(t))$ exhibits a monotonic upward trend over time $t$, smoothly approaching the upper bound value  1. This means that the quantum imaginarity is increasing with time, no decay phenomenon is observed. In addition, as $\phi$ increases, $\mathcal I^{G_2}(\rho(t))$ shows a more rapid ascent.  Fig.\ref{3}(b) demonstrates that at   given time $t=1,2,3$, $\mathcal I^{G_2}(\rho(t))$ exhibits an oscillatory behavior as a function of the parameter $\phi$; and as $t$ increases, the amplitude of these oscillations grows.

\begin{figure*}[tbp]
\center
\subfigure []
{\includegraphics[width=8cm,height=6cm]{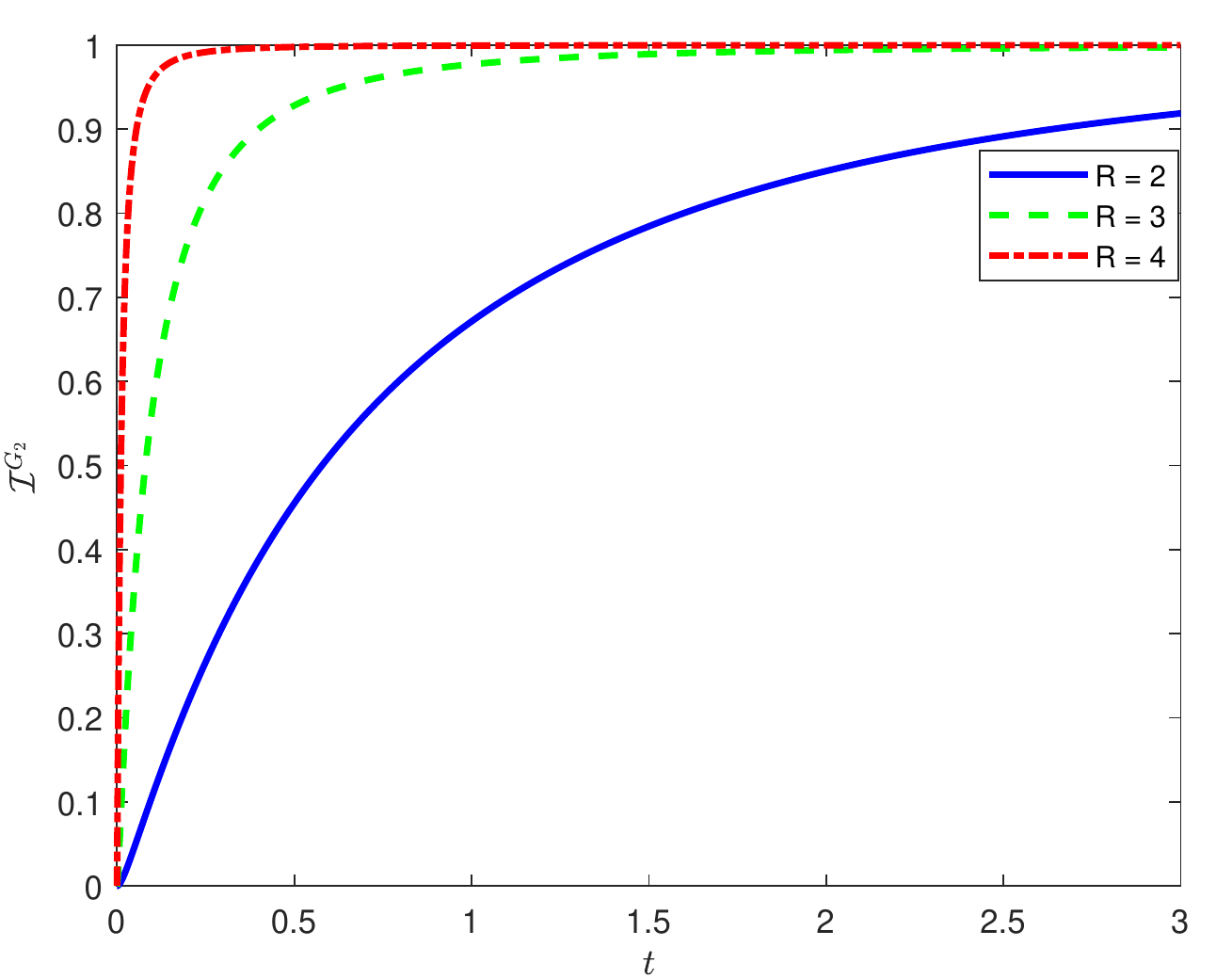}}
\subfigure []
{\includegraphics[width=8cm,height=6cm]{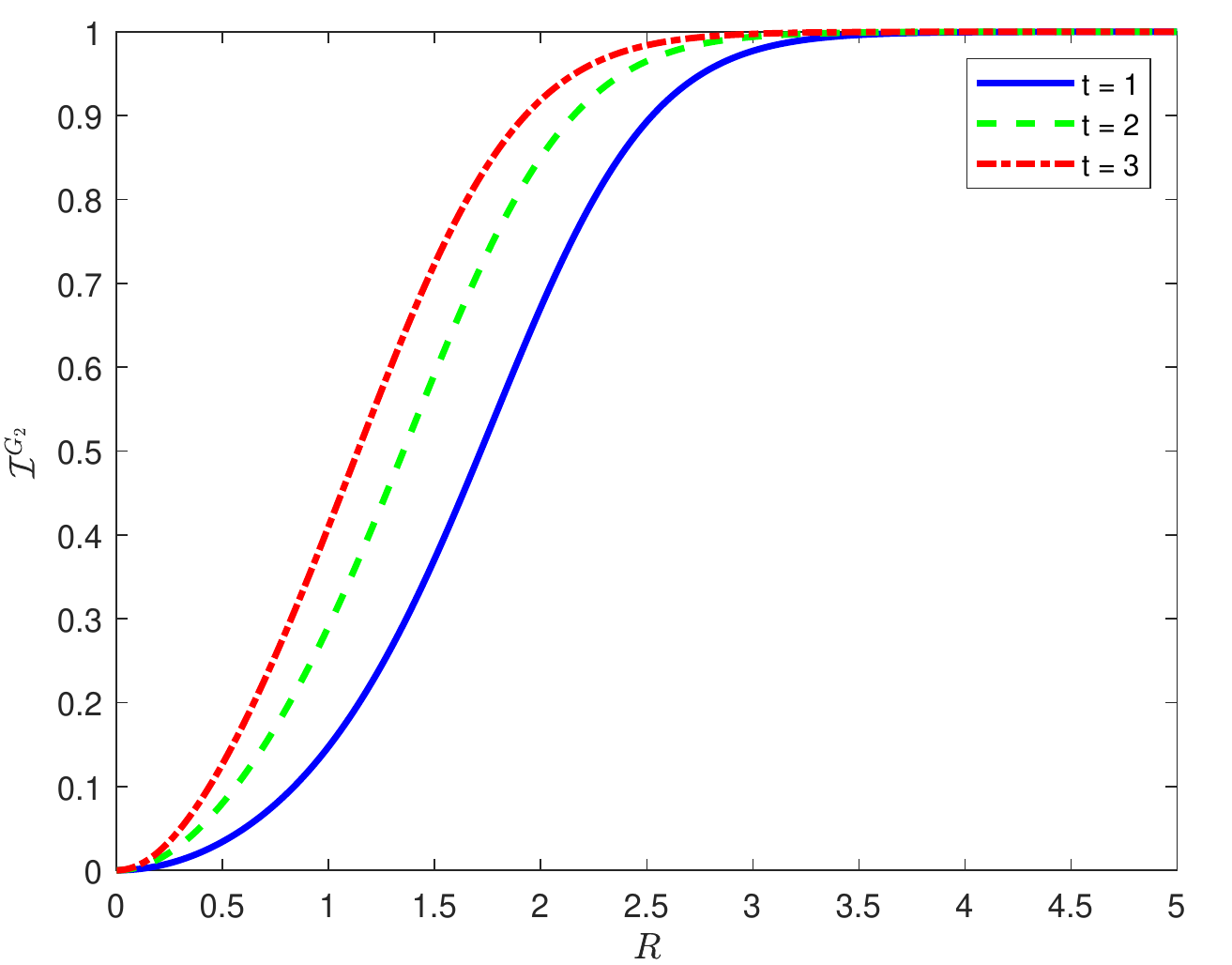}}
  \caption{\small Behavior of $\mathcal I^{G_2}(\rho(t))$, with $\rho(0)$ the squeezed vacuum state of $r=1$, as a function of the parameters $t$ and $R$ for fixed $n_\mathrm{th}=1.5, \phi=\frac{\pi}{2}, \lambda=0.1$. (a) Plots of $\mathcal I^{G_2}$ as a function of the parameter $t$ for fixed $R= 2, 3,4$, respectively.  (b) Plots of $\mathcal I^{G_2}$ as a function of the parameter $R$ for fixed $t=1,2,3$, respectively.}
\label{4}
\end{figure*}

In Fig.\ref{4},  we fix the parameters $r$, $n_\mathrm{th}$, $\phi$, $\lambda$ in $\nu_{\rho(t)}$ and take $r=1$, $n_\mathrm{th}=1.5, \phi=\pi/2, \lambda=0.1$. Thus,  $\mathcal I^{G_2}(\rho(t))$ is a function of $t$ and $R$. Fig.\ref{4} illustrates how $\mathcal I^{G_2}(\rho(t))$ varies with parameters $t$ and $R$. In Fig. \ref{4}(a), it can be observed that $\mathcal I^{G_2}(\rho(t))$ accelerates its approach towards 1 as $R$ increases. Additionally, Fig.\ref{4}(b) reveals that, for a fixed $t$,  $\mathcal I^{G_2}(\rho(t))$ increases with $R$,  until it approximates  to 1.

\begin{figure*}[tbp]
\center
\subfigure []
{\includegraphics[width=8cm,height=6cm]{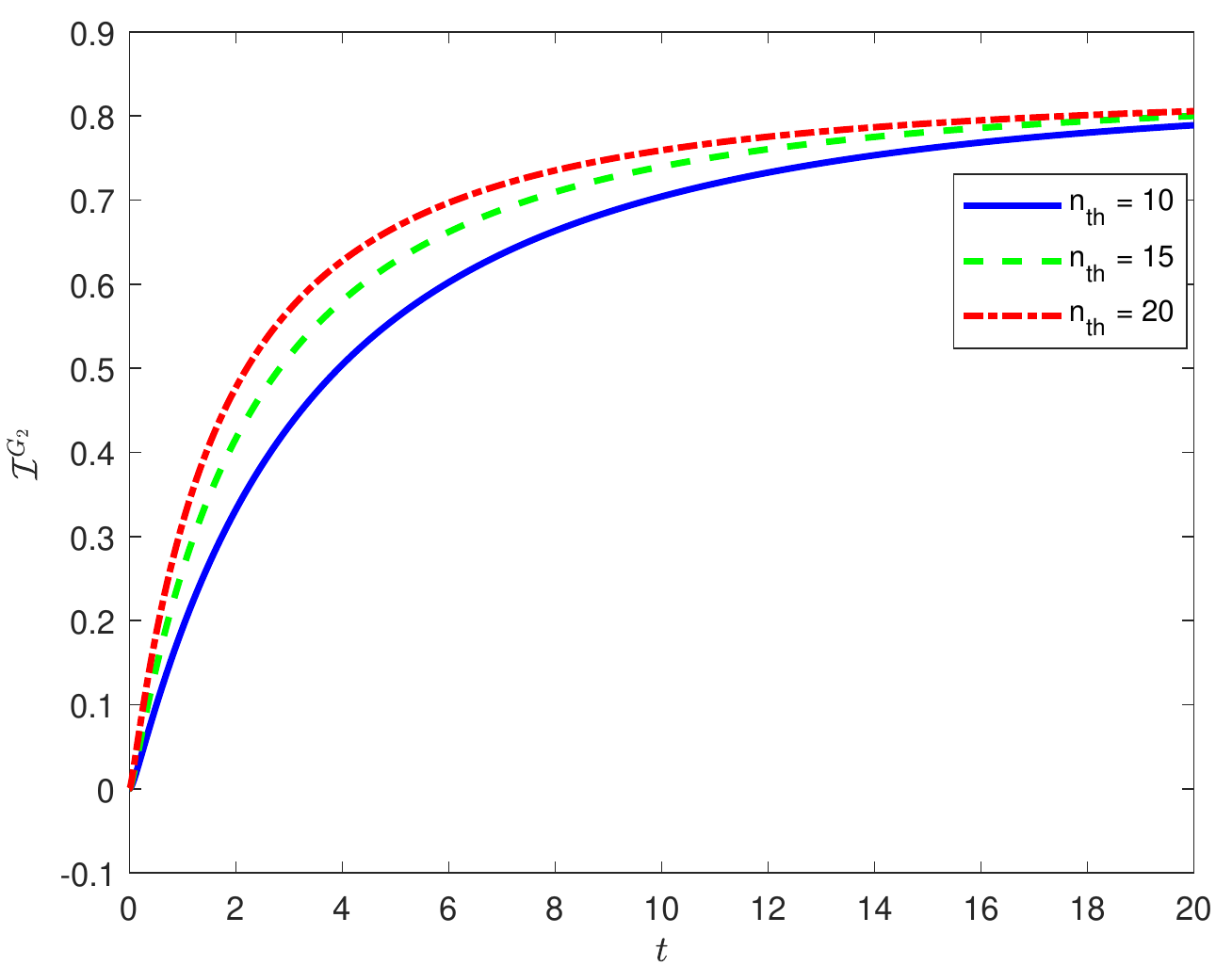}}
\subfigure []
{\includegraphics[width=8cm,height=6cm]{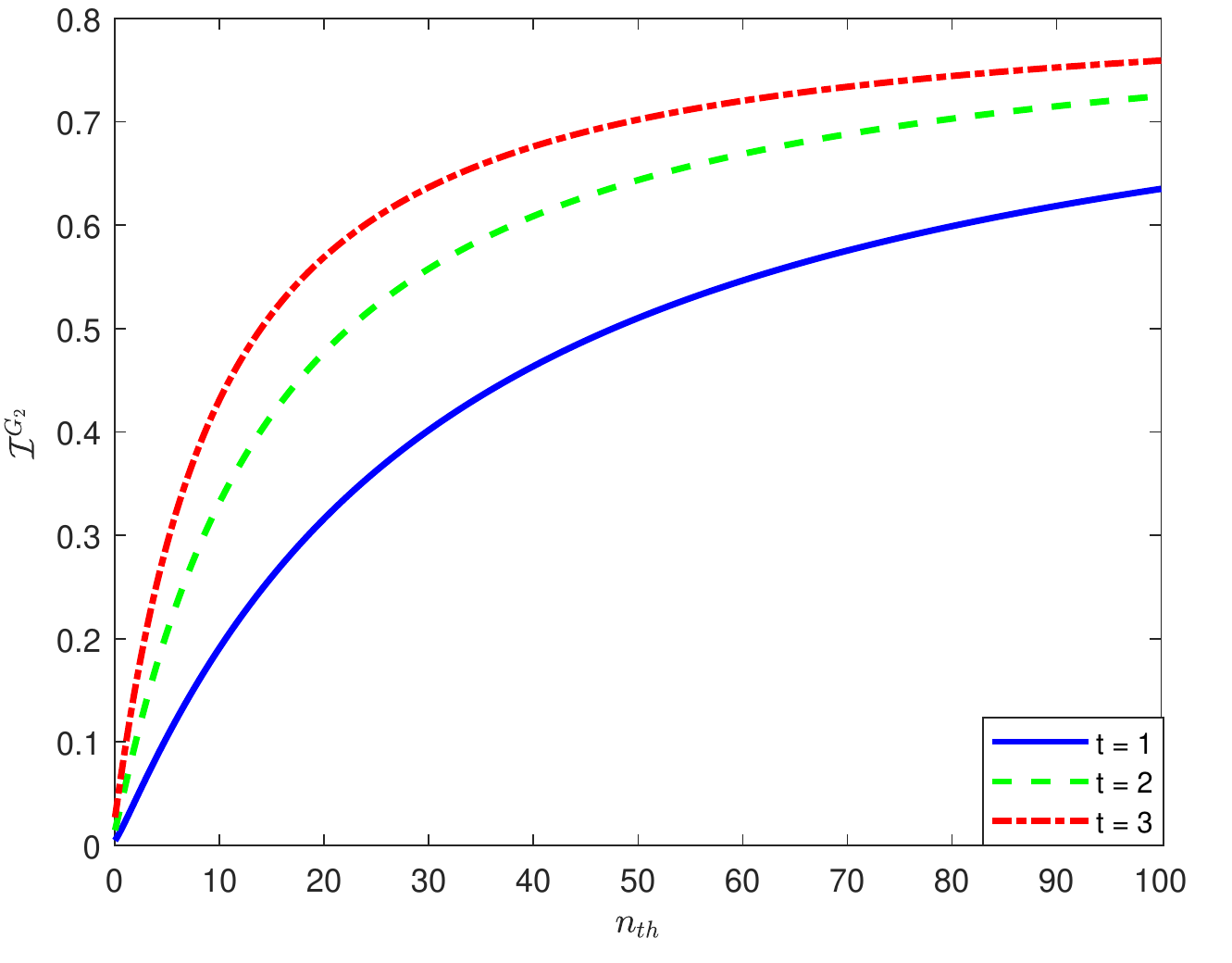}}
  \caption{Behavior of $\mathcal I^{G_2}(\rho(t))$, with $\rho(0)$ the squeezed vacuum state of $r=1$, as a function of the parameters $t$ and $n_\mathrm{th}$ for fixed $R=1, \phi=\frac{\pi}{2}, \lambda=0.1$. (a) plot of $\mathcal I^{G_2}$ as a function of the parameter $t$ for fixed $n_\mathrm{th}= 10, 15, \ {\rm and} \ 20$ and (b) plot of $\mathcal I^{G_2}$ as a function of the parameter $n_\mathrm{th}$ for fixed $t=1,2, \ {\rm and} \ 3$.}
\label{5}
\end{figure*}

Fig.\ref{5} illustrates the behavior of $\mathcal I^{G_2}(\rho(t))$ as a function of the parameters $t$ and $n_\mathrm{th}$ for fixed \textcolor{red}{$r=1$,} $R=1$, $\phi=\pi/2$  and $\lambda=0.1$. From Fig.\ref{5}(a), it is evident that as $n_\mathrm{th}$ increases, the rate of increase in $\mathcal I^{G_2}(\rho(t))$ becomes more prominent. Similarly, Fig.\ref{5}(b) exhibits a comparable tendency when $\mathcal I^{G_2}(\rho(t))$ is considered as a function of $n_\mathrm{th}$.

Next, assume that the initial Gaussian state $\rho'(0)$ is a two-mode Glauber coherent state in Example 9, of which the CM is $\nu_{\rho'(0)}=I_4$ and displacement vector $\bar{d}_{\rho'(0)}=(2{\rm Re}\alpha_1, 2{\rm Im}\alpha_1, 2{\rm Re}\alpha_2, 2{\rm Im}\alpha_2)$ with $\alpha_1\notin \mathbb{R}$. Then $\rho'(0)$ is an imaginary state. We detect the Gaussian imaginarity  $\mathcal I^{G_2}(\rho'(t))$. By Eqs.\eqref{eq20}-\eqref{eq21} again, we get
$$\nu_{\rho'(t)}=\left(\begin{array}{cccc}
a_{+}&  c & 0 & 0\\
c &a_{-}& 0 & 0\\
0 & 0 & a_{+} &c\\
0 & 0 & c &a_{-}\end{array}\right),$$
where $a_{\pm}=e^{-\lambda t}+(1-e^{-\lambda t})(1+2L_{\pm})$ and $c=2(1-e^{-\lambda t}){\rm Im}M$. Also, by Eq.\eqref{eq22},
$$\bar{d}_{\rho'(t)}=2e^{-\frac{\lambda t}{2}}({\rm Re}\alpha_1, {\rm Im}\alpha_1, {\rm Re}\alpha_2, {\rm Im}\alpha_2).$$
Now, by Eq.\eqref{eq12}, we obtain
\begin{equation*}\label{eq24}
\begin{aligned}
\mathcal I^{G_2}(\rho'(t))&=1-\frac{\det(\nu_{\rho'(t)})}{a_{+}^2a_{-}^2}\\
&+h(|2e^{-\frac{\lambda t}{2}}({\rm Im}\alpha_1)|+|2e^{-\frac{\lambda t}{2}}({\rm Im}\alpha_2)|)\\
&=2-\frac{(a_{+}a_{-}-c^2)^2}{a_{+}^2a_{-}^2}.
\end{aligned}
\end{equation*}

\begin{figure*}[tbp]
\center
\subfigure []
{\includegraphics[width=8cm,height=6cm]{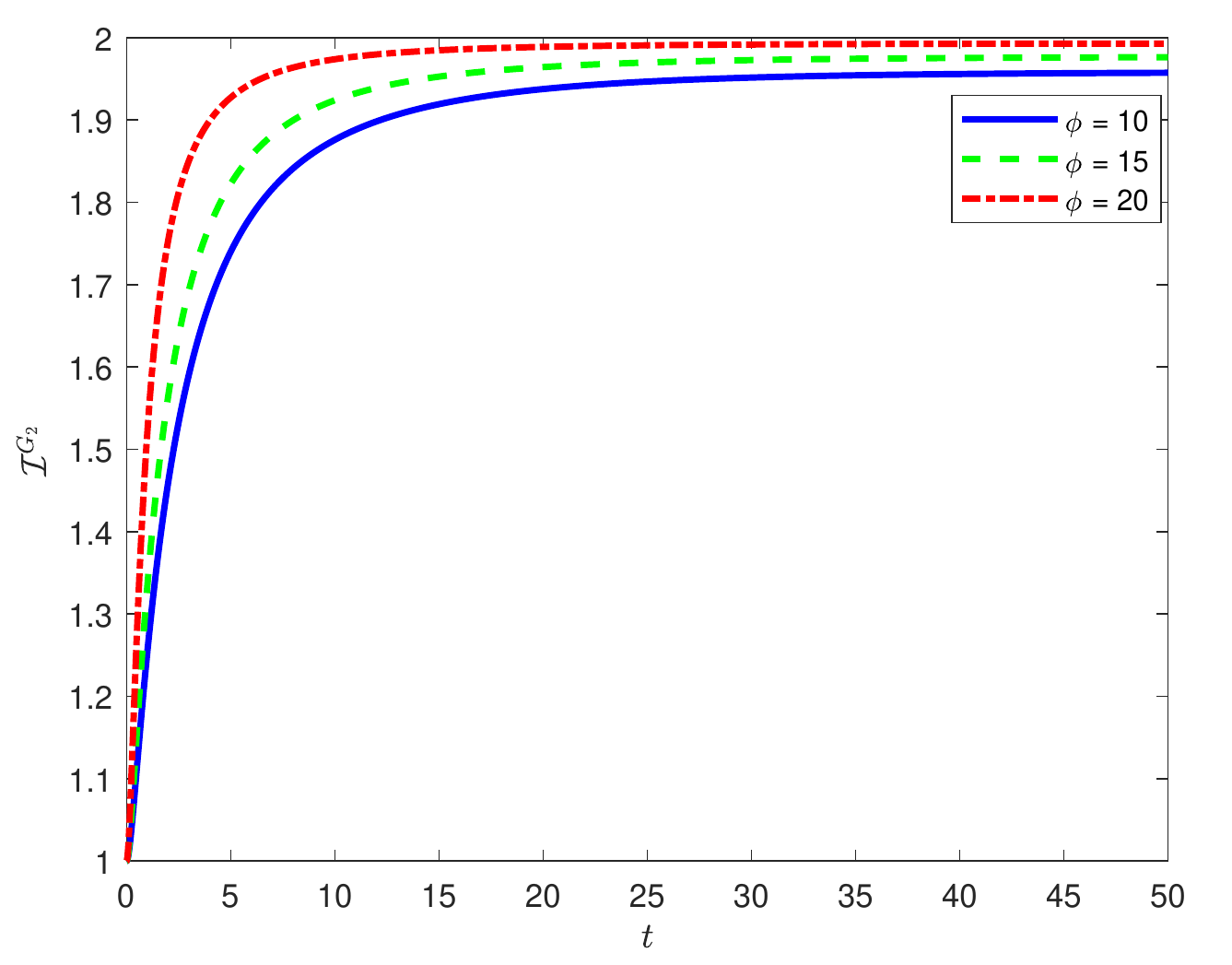}}
\subfigure []
{\includegraphics[width=8cm,height=6cm]{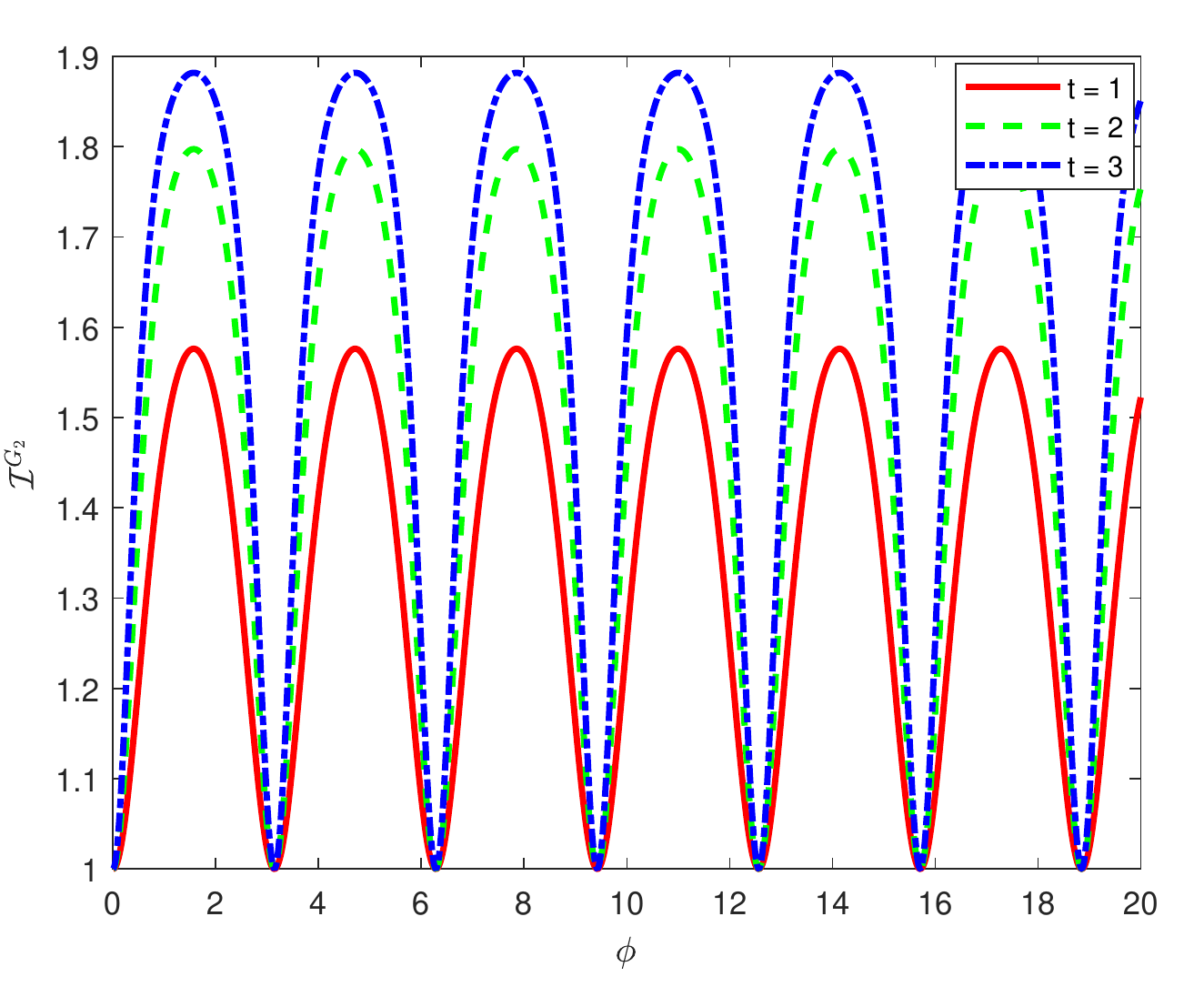}}
  \caption{\small Behavior of $\mathcal I^{G_2}(\rho'(t))$, with $\rho'(0)$ the Glauber coherent state of $r=1$, as a function of the parameters $t$ and $\phi$ for fixed $n_\mathrm{th}=1.5, R = 1, \lambda=0.1$. (a) plots of $\mathcal I^{G_2}$ as a function of the parameter $t$ for fixed $\phi= 10, 15, \ {\rm and} \ 20$ and (b) plots of $\mathcal I^{G_2}$ as a function of the parameter $\phi$ for fixed $t=1,2, \ {\rm and} \ 3$.}
\label{6}
\end{figure*}

\begin{figure*}[tbp]
\center
\subfigure []
{\includegraphics[width=8cm,height=6cm]{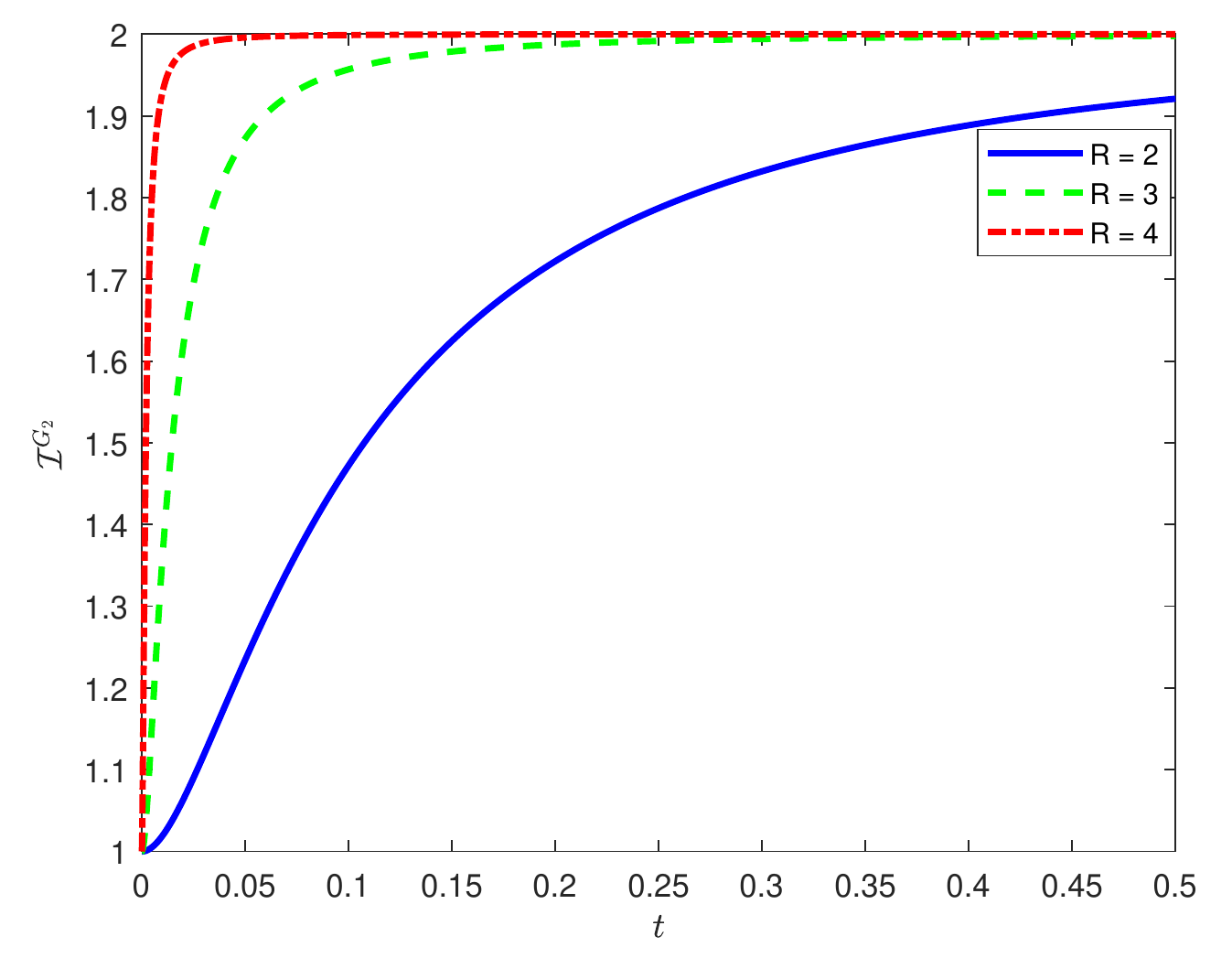}}
\subfigure []
{\includegraphics[width=8cm,height=6cm]{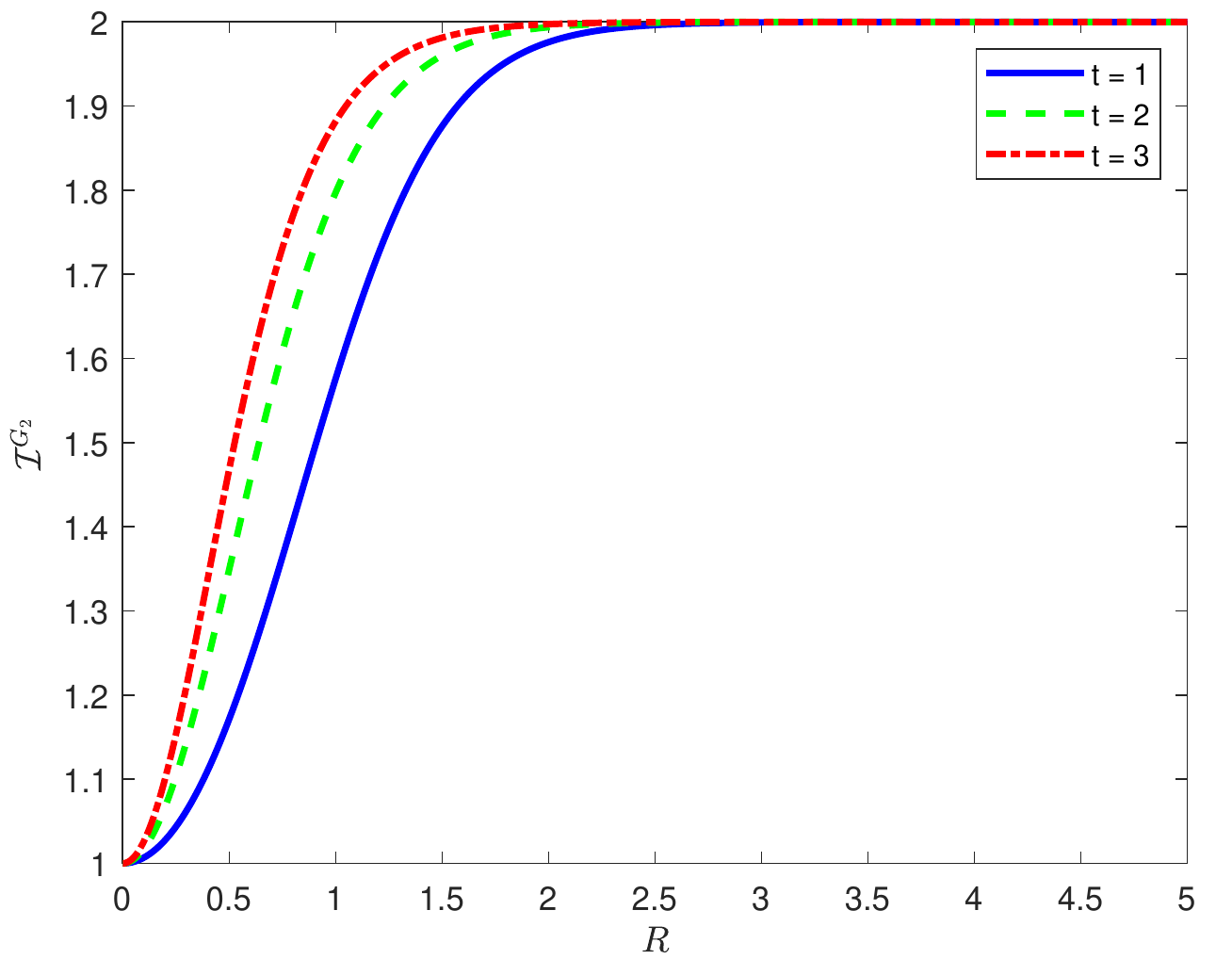}}
  \caption{\small Behavior of $\mathcal I^{G_2}(\rho'(t))$, with $\rho'(0)$ the Glauber coherent state of $r=1$, as a function of the parameters $t$ and $R$ for fixed $n_\mathrm{th}=1.5, \phi=\frac{\pi}{2}, \lambda=0.1$. (a) plot of $\mathcal I^{G_2}$ as a function of the parameter $t$ for fixed $R= 2, 3, \ {\rm and} \ 4$ and (b) plot of $\mathcal I^{G_2}$ as a function of the parameter $R$ for fixed $t=1,2, \ {\rm and} \ 3$.}
\label{7}
\end{figure*}

\begin{figure*}[tbp]
\center
\subfigure []
{\includegraphics[width=8cm,height=6cm]{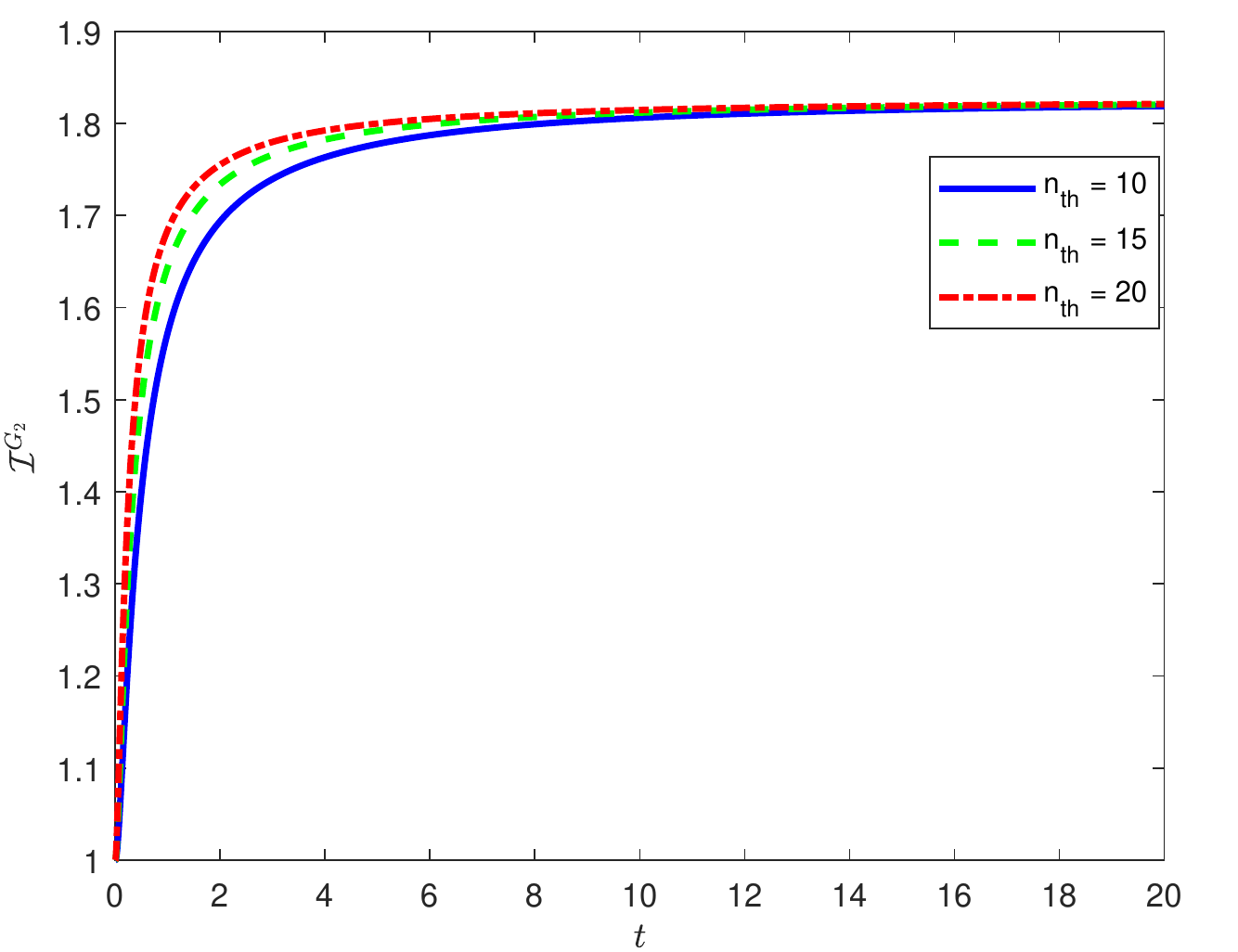}}
\subfigure []
{\includegraphics[width=8cm,height=6cm]{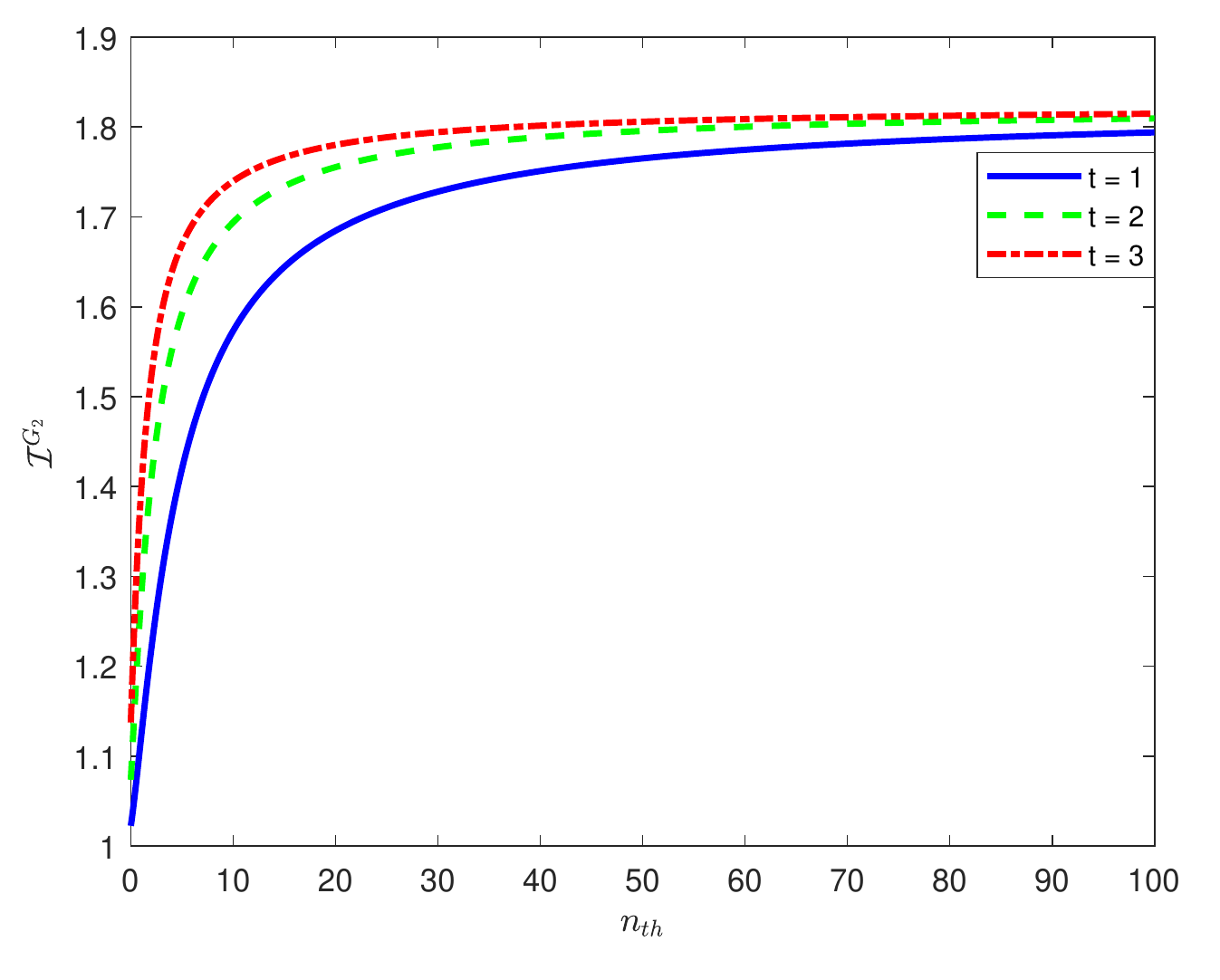}}
  \caption{\small Behavior of $\mathcal I^{G_2}(\rho'(t))$, with $\rho'(0)$ the Glauber coherent state of $r=1$, as a function of the parameters $t$ and $n_\mathrm{th}$ for fixed $R=1, \phi=\frac{\pi}{2}, \lambda=0.1$. (a) plots of $\mathcal I^{G_2}$ as a function of the parameter $t$ for fixed $n_\mathrm{th}= 10, 15, \ {\rm and} \ 20$ and (b) plots of $\mathcal I^{G_2}$ as a function of the parameter $n_\mathrm{th}$ for fixed $t=1,2, \ {\rm and} \ 3$.}
\label{8}
\end{figure*}

In Fig.\ref{6}-Fig.\ref{8}, when the initial Gaussian state is imaginarity state, we obtained a imaginarity dynamic evolution process similar to that in Fig.\ref{3}-Fig.\ref{5}.

\section{Multi-mode Gaussian imaginarity as a multipartite multi-mode Gaussian correlation}

Let $H_k$ be a $n_k$-mode CV system, $k=1,2,\ldots, m$. Then $H=H_1\otimes H_2\otimes\cdots\otimes H_m$ is an $m$-partite $n$-mode CV system with $n=n_H=(\sum_{k=1}^m n_k)$ the mode of $H$.  Regard $H$ as an $m$-partite system, an interesting question is to ask  whether or not the Gaussian quantum imaginarity is an $m$-partite Gaussian quantum correlation?

With respect to the product Fock basis of $H$, we have proved that ${\mathcal I}^{G_n}_{m}={\mathcal I}^{G_n}$ is a Gaussian imaginarity measure of CV system $H$. If we can show that ${\mathcal I}_m^{G_n}$ satisfies the requirements of an $m$-partite Gaussian correlation measures, then the Gaussian quantum imaginarity of CV system $H$ can induce an $m$-partite Gaussian correlation. Furthermore, note that, for  any Gaussian channel $\Phi_k$ of $H_k$, the local Gaussian channel $\Phi=\Phi_1\otimes\Phi_2\otimes\cdots\otimes \Phi_m$ of $H$ is real if and only if either all $\Phi_k$s are completely real or all $\Phi_k$s are covariant real. If ${\mathcal I}_m^{G_{n}}$ is a multipartite Gaussian correlation measure, then, with real Gaussian states as free states and real local Gaussian channels as free operations, the multipartite multi-mode Gaussian imaginarity is a multipartite quantum resource.

By Theorems \ref{thm5}-\ref{thm6},  it is easily checked that

(GI1)  ${\mathcal I}_m^{G_n}(\rho)\geq0$ for any Gaussian state $\rho\in{\mathcal S}(H_1\otimes H_2\otimes\cdots\otimes H_m)$, and ${\mathcal I}_m^{G_n}(\rho)=0$ if and only if $\rho$ is real;

(GI2) ${\mathcal I}_m^{G_n}(\Phi(\rho))\leq {\mathcal I}_m^{G_n}(\rho)$ for any Gaussian state $\rho\in{\mathcal S}(H_1\otimes H_2\otimes\cdots\otimes H_m)$ and any real local Gaussian channel $\Phi=\Phi_1\otimes\Phi_2\otimes\cdots\otimes \Phi_m$.

Also note that the  ${\mathcal I}_m^{G_n}$ is invariant under  order changes of subsystems, which reflects the symmetry of Gaussian quantum imaginarity with respect to subsystems.

However, as an $m$-partite quantum correlation measure, by \cite{HLQ22}, ${\mathcal I}_m^{G_n}$ should also satisfy the unification condition and the hierarchy condition. Thus, we should consider not only ${\mathcal I}_m^{G_n}$ but also the set $\{ {\mathcal I}_k^{G_r}: 2\leq r\leq n, 2\leq k\leq \min\{r,m\} \}$. The unification condition is obviously satisfied as each ${\mathcal I}_k^{G_r}$ is defined in the same way as ${\mathcal I}_m^{G_n}$. The hierarchy condition is a requirement coming from resource allocation theory, which ensures that the imaginarity contained in the part systems does not exceed the imaginarity hold by the whole system. By \cite{HLQ22}, to verify the hierarchy condition for ${\mathcal I}_m^{G_n}$, one has to prove the following result.

\begin{theorem}\label{thm10}
${\mathcal I}_m^{G_{n_H}}$ satisfies the hierarchy condition, that is,

{\rm (1)} for $2\leq r<m$ and any Gaussian state $\rho=\rho_{1,2,\ldots,m}\in{\mathcal S}(H_1\otimes H_2\otimes\cdots\otimes H_m)$,  denote by $H=H_1\otimes H_2\otimes\cdots\otimes H_m$ and $K=H_1\otimes H_2\otimes\cdots\otimes H_r$, we have ${\mathcal I}_r^{G_{n_K}}(\rho_K)\leq {\mathcal I}_m^{G_{n_H}}(\rho)$, where $\rho_K={\rm Tr}_{K^c}(\rho)$ is the reduced state of $\rho$ with respect to the subsystem $K$;

{\rm (2)} for any $r$-partition $P=P_1|P_2|\ldots|P_r$ of $\{1,2,\ldots,m\}$ and any Gaussian state $\rho\in{\mathcal S}(H_1\otimes H_2\otimes \cdots\otimes H_m)$, regarding also $\rho\in {\mathcal S}(H_{P_1}\otimes H_{P_2}\otimes\cdots\otimes H_{P_r})$ as an $m$-partite Gaussian state, we have ${\mathcal I}_r^{G_{n_H}}(\rho)\leq {\mathcal I}_m^{G_{n_H}}(\rho)$;

{\rm (3)} for any $r$-partition $P=P_1|P_2|\ldots|P_r$ of $\{1,2,\ldots,m\}$, any $Q=Q_1|Q_2|\ldots|Q_r$ obtained by $\emptyset\not= Q_j\subseteq P_j$, $j=1,2,\ldots,r$, and for any Gaussian state $\rho\in{\mathcal S}(H_1\otimes H_2\otimes\cdots\otimes H_m)$, regarding $\rho\in{\mathcal S}(H_{P_1}\otimes H_{P_2}\otimes\cdots\otimes H_{P_r})$ and $\rho_Q\in{\mathcal S}(H_{Q_1}\otimes H_{Q_2}\otimes\cdots\otimes H_{Q_r})$ as $m$-partite Gaussian states, where $\rho_Q$ is the reduced state of $\rho$ with respect to the subsystem $H_Q=H_{Q_1}\otimes H_{Q_2}\otimes\cdots\otimes H_{Q_r}$, we have ${\mathcal I}_r^{G_{n_{H_Q}}}(\rho_Q)\leq {\mathcal I}_r^{G_{n_H}}(\rho)$.
\end{theorem}

A proof  is presented in  the Appendix.

Therefore, ${\mathcal I}_m^{G_{n}}$ meets all requirements as a multipartite multi-mode Gaussian correlation measures.  This unveils that  the Gaussian quantum imaginarity can also be regarded as an $m$-partite multi-mode Gaussian quantum correlation, and, with measure ${\mathcal I}_m^{G_n}$, is a multipartite Gaussian quantum resource.

\section{Conclusions}

This work establishes an accessible quantification scheme $\mathcal{I}^{G_n}$ for Gaussian quantum imaginarity in arbitrary $n$-mode continuous-variable systems, which is uniquely determined by the displacement vectors and covariance matrices of Gaussian states, requiring no additional parameters.  $\mathcal{I}^{G_n}$ satisfies the properties (i) faithfulness and (ii) monotonicity under Gaussian real channels, makes it a valid multi-mode measure of  Gaussian quantum imaginarity. It is interesting to note that, by $\mathcal{I}^{G_1}$,  the single-mode Gaussian imaginarity is independent of the average photon number
of Gaussian states.

Compared   $\mathcal{I}^{G_n}$ with existing Gaussian imaginarity measures, particularly the fidelity-based Gaussian imaginarity measure $M_F$ and the Tsallis relative entropy Gaussian imaginarity measure $M_{T,\mu}$, the key advantages of $\mathcal{I}^{G_n}$ manifest in computational efficiency, which makes $\mathcal{I}^{G_n}$ particularly suitable in the scenarios of quantum information tasks utilizing Gaussian imaginarity. As an application of ${\mathcal I}^{G_n}$,  we study the dynamics behaviour of Gaussian imaginarity  in Gaussian (Markovian) noise environments for $2$-mode CV systems. By considering respectively the squeezed vacuum state (a real state) and  the Glauber coherent state (an imaginary Gaussian state) as initial Gaussian states,  we see that ${\mathcal I}^{G_n}$ increases as time goes by and tends to the upper bound rapidly.

Though the Gaussian imaginarity is not a Gaussian quantum correlation as it lives essentially in single-partite systems, we also study the question  whether multi-mode Gaussian imaginarity can be regarded as a kind of multipartite multi-mode Gaussian correlation.  This question is attacked and solved by utilizing ${\mathcal I}^{G_n}$ as it induces an $m$-partite multi-mode Gaussian correlation measure ${\mathcal I}_m^{G_n}$ which, besides the faithfulness and nonincreasing trend under local real Gaussian operations,  satisfies further the symmetry with respect to subsystems, the unification condition and the hierarchy condition. This further demonstrates that the Gaussian quantum imaginarity can be also regarded as a multipartite Gaussian quantum resource.

 \vspace{2mm}
{\bf Acknowledgments}
This work is supported by National Natural
Foundation of China (12301152, 12171290, 12071336) and the Natural Science Foundation of Shanxi Province (202203021222018).

\section*{Appendix}  
\renewcommand{\theequation}{A\arabic{equation}}  
\setcounter{equation}{0}

To prove Theorems \ref{thm5} and \ref{thm6}, we need several lemmas.

\begin{lemma}\label{lem10} {\rm (\cite[Lemma A1]{HLQ22})} Assume that
$$\Gamma=\left(
           \begin{array}{cccc}
             \Gamma_{11} & \Gamma_{12} & \cdots & \Gamma_{1k} \\
             \Gamma_{21} & \Gamma_{22} & \cdots & \Gamma_{2k} \\
             \vdots & \vdots &  \ddots & \vdots \\
             \Gamma_{k1} & \Gamma_{k2} & \cdots & \Gamma_{kk} \\
           \end{array}
         \right)$$
is a positive definite block matrix over the complex field $\mathbb{C}$. Then $\det(\Gamma)=\prod^{k}_{i=1}\det(\Gamma_{ii})$ if and only if $\Gamma_{ij}=0$ whenever $i\ne j$.
\end{lemma}

\begin{lemma}\label{lem11} {\rm (\cite[Lemma A3]{HLQ22})}  Let $B,K,M\in\mathcal{M}_{n}(\mathbb{C})$ with $B$ and $M$ positive semi-definite.
If both $B$ and $KBK^{\dag}+M$ are invertible, then $K^{\dag}(KBK^{\dag}+M)^{-1}K\le B^{-1}$; and moreover, the equality holds if and only if $M=0$ and $K$ is invertible.
\end{lemma}

\begin{lemma}\label{lem12} {\rm (\cite{HJ})} The following statements hold.

{\rm (1)} Assume that $T=\left(
                                         \begin{array}{cc}
                                           A & C \\
                                           C^\dag & B \\
                                         \end{array}
                                       \right)$
is a block positive definite matrix. Then $\det(T)=\det(A)\det(B-C^\dag A^{-1}C)=\det(B)\det(A-CB^{-1}C^\dag)$.

{\rm (2)} For any positive definite matrices $A,B$, $A\le B$ implies $\det A\le \det B$.
\end{lemma}

{\bf Proof of Theorem \ref{thm5}.} $\mathcal{I}^{G_n}(\rho)\geq 0$ is obvious.  The sufficiency part  is evident by \cite[Theorem 1]{Xu1}. For the necessity part, assume that $\mathcal{I}^{G_n}(\rho)=0$.
By Eq.\eqref{eq12}, we have $$1+h(|d_2|+|d_4|+\cdots+|d_{2n}|)=\frac{\det(\nu)}{\det(A_{11})\det(A_{22})}.$$
Also note that $0<\det(\nu)=\det(P_n\nu P_n^{\rm T})\le\det(A_{11})\det(A_{22})$. The above equation implies
\begin{equation}\label{eqA1}
h(|d_2|+|d_4|+\cdots+|d_{2n}|)=0
\end{equation}
and
\begin{equation}\label{eqA2}
{\rm det}(P_n\nu P_n^{\rm T})={\rm det}(\nu)={\rm det}(A_{11}){\rm det}(A_{22}).\end{equation}
Eq.\eqref{eqA1} yields $d_2=d_4=\cdots=d_{2n}$,
and applying Lemma \ref{lem10} to Eq.\eqref{eqA2} gives $A_{12}=0$. So $\rho$ is real.\hfill$\Box$

{\bf Proof of Theorem \ref{thm6}.} Take any $n$-mode Gaussian state $\rho=\rho(\bar{d},\nu)$  and any real Gaussian channel $\Phi$.
If $\Phi$ is a completely real Gaussian channel,
by \cite[Theorem 4]{Xu1}, $\Phi(\rho)$ is real. As a result, we have $\mathcal{I}^{G_n}(\Phi(\rho))=0\le\mathcal{I}^{G_n}(\rho)$.
Now, assume that $\Phi$ is a covariant real Gaussian channel. In this case, $\Phi$ is determined by the map $\bar{d}\mapsto\bar{d}'=\bar{d}_{\Phi(\rho)}=T\bar{d}+\bar{d}_0$,
$\nu\mapsto\nu'=\nu_{\Phi(\rho)}=T\nu T^{\rm T}+N$. Here $\bar{d}_0=(d^0_1,0,d^0_3,0,\ldots,d^0_{2n-1},0)^{\rm T}\in\mathbb{R}^{2n}$,
$T=(t_{ij})$ and $N=N^{\rm T}=(n_{ij})\ge0$ are $2n\times 2n$ real matrices which satisfy the following conditions:
$$t_{2k-1,2l}=t_{2k,2l-1}=0 \ \ {\rm for} \ \ k,l\in\{1,2,\ldots,n\}$$
and $$n_{2k-1,2l}=0 \ \ {\rm for} \ \ k,l\in\{1,2,\ldots,n\}.$$
Consequently,  $P_n\bar{d_0}=(d',0)^{\rm T}$,
$P_nTP_n^{\rm T}=\left(
                  \begin{array}{cc}
                    T_1 & 0 \\
                    0 & T_2 \\
                  \end{array}
                \right)$
and $P_nNP_n^{\rm T}=\left(
                  \begin{array}{cc}
                    N_1 & 0 \\
                    0 & N_2 \\
                  \end{array}
                \right),$
where $d'\in{\mathbb R}^{n}$, $T_1, T_2, N_1, N_2\in\mathcal{M}_{n}(\mathbb{R})$. Therefore,
\begin{equation}\label{eqA3}
\begin{array}{rl}Q_n'P_n\bar{d}'= & Q_n'P_n(T\bar{d}+\bar{d_0})\\
= & Q_n'(P_nTP_n^{\rm T})(P_n\bar{d})+Q_n'P_n\bar{d_0}\\
= & Q_n'\left(
                  \begin{array}{cc}
                    T_1 & 0 \\
                    0 & T_2 \\
                  \end{array}
                \right)\left(
                         \begin{array}{c}
                           Q_nP_n\bar{d} \\
                           Q_n'P_n\bar{d} \\
                         \end{array}
                       \right)\\
= & T_2Q_n'P_n\bar{d}\end{array}\end{equation}
and

\begin{equation}\label{eqA4}
\begin{array}{rl} &P_n\nu'P_n^{\rm T}\\= &P_n(T\nu T^{\rm T}+N)P_n^{\rm T}\\
= & (P_nTP_n^{\rm T})(P_n\nu P_n^{\rm T})(P_nT^{\rm T}P_n^{\rm T})+P_nNP_n^{\rm T}\\
= & \left(\begin{array}{cc}
                 T_1A_{11}T_1^{\rm T}+N_1     & T_1A_{12}T_2^{\rm T} \\
                 T_2A_{12}^{\rm T}T_1^{\rm T} & T_2A_{22}T_2^{\rm T}+N_2\\
                  \end{array}
                \right).
\end{array}\end{equation}

By Eq.\eqref{eq11}, together with Eqs.\eqref{eqA3}-\eqref{eqA4}, we can derive that
$$\begin{array}{rl}&\mathcal{I}^{G_n}(\Phi(\rho))\\
= & 1-\dfrac{\det(\nu')}{\det(Q_nP_n\nu'P_n^{\rm T}Q_n^{\rm T})\det(Q'_nP_n\nu'P_n^{\rm T}Q_n'^{\rm T})}\\&+h(\|Q_n'P_n\bar{d}'\|_1)\\
= & 1-\dfrac{\det(P_n\nu'P_n^{\rm T})}{\det(T_1A_{11}T_1^{\rm T}+N_1)\det(T_2A_{22}T_2^{\rm T}+N_2)}\\&+h(\|T_2Q_n'P_n\bar{d}\|_1)\\
\le & 1-\dfrac{\det(P_n\nu'P_n^{\rm T})}{\det(T_1A_{11}T_1^{\rm T}+N_1)\det(T_2A_{22}T_2^{\rm T}+N_2)}\\&+h(\|Q_n'P_n\bar{d}\|_1).
\end{array}$$

Thus, to prove $\mathcal{I}^{G_n}(\Phi(\rho))\le\mathcal{I}^{G_n}(\rho)$, it suffices to show that
$$\begin{array}{rl}&\dfrac{\det(P_n\nu'P_n^{\rm T})}{\det(T_1A_{11}T_1^{\rm T}+N_1)\det(T_2A_{22}T_2^{\rm T}+N_2)}\\
	\ge&\dfrac{\det(P_n\nu P_n^{\rm T})}{\det(A_{11})\det(A_{22})}.\end{array}$$
In fact, by using Lemma \ref{lem11} twice and Lemma \ref{lem12}, one has

\begin{widetext}
$$\begin{array}{rl}
& \dfrac{\det(P_n\nu'P_n^{\rm T})}{\det(T_1A_{11}T_1^{\rm T}+N_1)\det(T_2A_{22}T_2^{\rm T}+N_2)}\\
= & \dfrac{\det(T_1A_{11}T_1^{\rm T}+N_1)\det[(T_2A_{22}T_2^{\rm T}+N_2)-(T_2A_{12}^{\rm T}T_1^{\rm T})(T_1A_{11}T_1^{\rm T}+N_1)^{-1}T_1A_{12}T_2^{\rm T}]}{\det(T_1A_{11}T_1^{\rm T}+N_1)\det(T_2A_{22}T_2^{\rm T}+N_2)}\\
\ge &\dfrac{\det[(T_2A_{22}T_2^{\rm T}+N_2)-(T_2A_{12}^{\rm T}A_{11}^{-1}A_{12}T_2^{\rm T})]}{\det(T_2A_{22}T_2^{\rm T}+N_2)}\\
= &\dfrac{\det(A_{11})\det[(T_2A_{22}T_2^{\rm T}+N_2)-(T_2A_{12}^{\rm T}A_{11}^{-1}A_{12}T_2^{\rm T})]}{\det(A_{11})\det(T_2A_{22}T_2^{\rm T}+N_2)}\
= \dfrac{\det\left(
                     \begin{array}{cc}
                       A_{11} & A_{12}T_2^{\rm T} \\
                       T_2A_{12}^{\rm T} & T_2A_{22}T_2^{\rm T}+N_2 \\
                     \end{array}
                   \right)}{\det(A_{11})\det(T_2A_{22}T_2^{\rm T}+N_2)}\\
= &\dfrac{\det(T_2A_{22}T_2^{\rm T}+N_2)\det[A_{11}-A_{12}T_2^{\rm T}(T_2A_{22}T_2^{\rm T}+N_2)^{-1}T_2A_{12}^{\rm T}]}{\det(A_{11})\det(T_2A_{22}T_2^{\rm T}+N_2)}\\
\ge &\dfrac{\det(A_{11}-A_{12}A_{22}^{-1}A_{12}^{\rm T})}{\det(A_{11})}
=\dfrac{\det(A_{22})\det(A_{11}-A_{12}A_{22}^{-1}A_{12}^{\rm T})}{\det(A_{22})\det(A_{11})}
= \dfrac{\det(P_n\nu P_n^{\rm T})}{\det(A_{11})\det(A_{22})}.
\end{array}$$
\end{widetext}
This completes the proof of Theorem \ref{thm6}.\hfill$\Box$

{\bf Proof of Eq.\eqref{eq13}.} For one-mode Gaussian state $\rho$, by Eq.\eqref{eq6}, $M_F(\rho)$ can be simplified to
$$\begin{array}{rl}M_F(\rho)=&1-\dfrac{\exp(-\frac{1}{4}(\bar d-\bar {d^*})^{\rm T}(\nu+\nu^*)^{-1}(\bar d-\bar {d^*}))}{\sqrt{\sqrt{\det(\frac{\nu+\nu^*}{2})+\Lambda}-\sqrt{\Lambda}}}\\
=&1-\dfrac{\exp(-\frac{d_2^2}{2\nu_{22}})}{\sqrt{\sqrt{\nu_{11}\nu_{22}+\Lambda}-\sqrt{\Lambda}}},\end{array}$$
with
$$\begin{array}{rl}\Lambda=&4\det\left(\frac {\nu + i\Delta_1}{2}\right)\det\left(\frac {\nu^*+i\Delta_1}{2}\right)\\
=&\dfrac{(\nu_{11}\nu_{22}-\nu_{12}^2-1)^2}{4}\end{array}.$$
Now, according to the displacement vector and CM of $\rho$, Eq.\eqref{eq13} is proved.\hfill$\Box$

{\bf Proof of Eq.\eqref{eq14}.}
For one-mode Gaussian state $\rho$, by Eq.\eqref{eq6}, $M_F(\rho)$ can be simplified to
\begin{equation}\label{eq141}\begin{aligned} &M_{T,\mu}(\rho)\\&=1-\frac{2(1-e^{-\eta_1})}{(1-e^{-\mu\eta_1})(1-e^{-(1-\mu)\eta_1})\sqrt{\det(\nu^{(\mu)}+\nu^{*(1 -\mu)})}}\\
&\times\exp\left\{-\frac{1}{2}(\bar{d}-\bar{d}^*)^{\rm T}(\nu^{(\mu)}+\nu^{*(1-\mu)})^{-1}(\bar{d}-\bar{d}^*)\right\}.
\end{aligned}\end{equation}

By the fact $\det(\Delta_1\nu+i\nu_1I_2)=0$, it can be deduced that $\nu_1=(1+2n_\mathrm{th})$. So $\eta_1=\ln\frac{n_\mathrm{th}+1}{n_\mathrm{th}}$.
For any $\mu$ with $0<\mu<1$, this implies
\begin{equation}\label{eq142}
\begin{cases}
e^{-\eta_1}=x, \ e^{-\mu\eta_l}=x^{\mu},\\
e^{-(1- \mu)\eta_l}=x^{(1-\mu)},
\end{cases}\end{equation}
with $x=\frac{n_\mathrm{th}}{n_\mathrm{th}+1}$.
By $\nu$, we obtain
$$\nu^{-1}=\frac{1}{1+2{n_\mathrm{th}}}\begin{pmatrix}
a & -b \\
-b & c
\end{pmatrix}$$
where
$$\begin{cases}
a=\cosh(2|\zeta|)-\cos\theta\sinh(2|\zeta|),\\
b=\sin\theta\sinh(2|\zeta|),\\
c=\cosh(2|\zeta|)+\cos\theta\sinh(2|\zeta|),\end{cases}$$
and
$$\begin{aligned}&\nu^{-\frac{1}{2}}=\frac{1}{\sqrt{1+2{n_\mathrm{th}}}}\\
&\times\begin{pmatrix}
\cosh(|\zeta|)-\cos\theta\sinh(|\zeta|) & -\sin\theta\sinh(|\zeta|) \\
-\sin\theta\sinh(|\zeta|) & \cosh(|\zeta|)+\cos\theta\sinh(|\zeta|)
\end{pmatrix}.\end{aligned}$$
Let $D=\mathrm{diag}(\nu_1,\nu_1)$, it follows from $S=D^{-\frac{1}{2}}\Delta_1\nu^{-\frac{1}{2}}$ that
$$S =\begin{pmatrix}
    \cosh(|\zeta|) - \cos\theta \sinh(|\zeta|) & -\sin\theta \sinh(|\zeta|) \\
    \sin\theta \sinh(|\zeta|) & -\cosh(|\zeta|) - \cos\theta \sinh(|\zeta|)
\end{pmatrix}.$$
According to $\nu^{(\mu)}=\nu_1^{(\mu)}SS^{\rm T}$ and $\nu^{*(1-\mu)}=\nu_1^{(1-\mu)}O_1SS^{\rm T}O_1$, we deduce that
$$\begin{cases}\nu^{(\mu)}
=\nu^{(\mu)}_1\begin{pmatrix}
    a & b \\
    b & c
\end{pmatrix},\\
\nu^{*(1-\mu)}
=\nu^{(1-\mu)}_1\begin{pmatrix}
    a & -b\\
   -b & c
\end{pmatrix}.\end{cases}$$
So,
\begin{equation}\label{eq143}
\det(\nu^{(\mu)}+\nu^{*(1-\mu)})=(\nu^{(\mu)}_1+\nu^{(1-\mu)}_1)^2
+4\nu^{(\mu)}_1\nu^{(1-\mu)}_1b^2\end{equation}
and
\begin{equation}\label{eq144}\begin{aligned}
&(\nu^{(\mu)}+\nu^{*(1-\mu)})^{-1}\\
&=\frac{1}{(\nu^{(\mu)}_1+\nu^{(1-\mu)}_1)^2
+4\nu^{(\mu)}_1\nu^{(1-\mu)}_1b^2}\\
&\times\begin{pmatrix}
(\nu^{(\mu)}_1+\nu^{(1-\mu)}_1)c & -(\nu^{(\mu)}_1-\nu^{(1-\mu)}_1)b\\
-(\nu^{(\mu)}_1-\nu^{(1-\mu)}_1)b & (\nu^{(\mu)}_1+\nu^{(1-\mu)}_1)a
\end{pmatrix}.\end{aligned}\end{equation}
Also, $\nu_1$ yields
\begin{equation}\label{eq145}
\begin{cases}\nu^{(\mu)}_1=\dfrac{2}{1-(\frac{n_\mathrm{th}}{n_\mathrm{th}+1})^{\mu}}-1,\\
\nu^{(1-\mu)}_1=\dfrac{2}{1-(\frac{n_\mathrm{th}}{n_\mathrm{th}+1})^{(1-\mu)}}-1.
\end{cases}\end{equation}
Moreover, $\bar{d^*}=O_1\bar d$ yields that $\bar d-\bar{d^*}=(0,4{\rm Im}\alpha)^{\rm T}$. Substitute these along with Eqs. \eqref{eq142}-\eqref{eq145} into Eq.\eqref{eq141}, and then we can conclude that Eq.\eqref{eq14} holds.\hfill$\Box$

{\bf Calculations of $\bar{d}$ and $\nu$ in Example 10.} For any $1\le l\le n$, noting that $\langle\alpha_l|\hat a^\dag_l=\alpha_l^*\langle\alpha_l|$, we can derive that
$$\begin{array}{rl}\bar d_{2l-1}=&{\rm Tr}(\hat R_{2l-1}|\alpha\rangle\langle\alpha|)={\rm Tr}(\hat Q_l|\alpha\rangle\langle\alpha|)\\
	=&\langle\alpha_1\alpha_2\cdots\alpha_n|\hat Q_l|\alpha_1\alpha_2\cdots\alpha_n\rangle\\
=&\langle\alpha_1\alpha_2\cdots\alpha_n|(\hat a_l+\hat a^\dag_l)|\alpha_1\alpha_2\cdots\alpha_n\rangle\\
=&\langle\alpha_1\alpha_2\cdots\alpha_n|\hat a_l|\alpha_1\alpha_2\cdots\alpha_n\rangle\\&+\langle\alpha_1\alpha_2\cdots\alpha_n|\hat a^\dag_l|\alpha_1\alpha_2\cdots\alpha_n\rangle\\
=&\alpha_l+\alpha_l^*=2{\rm Re}\alpha_l
\end{array}$$
and
$$\begin{array}{rl}\bar d_{2l}=&{\rm Tr}(\hat R_{2l}|\alpha\rangle\langle\alpha|)={\rm Tr}(\hat P_l|\alpha\rangle\langle\alpha|)\\
	=&\langle\alpha_1\alpha_2\cdots\alpha_n|\hat P_l|\alpha_1\alpha_2\cdots\alpha_n\rangle\\
=&-i(\langle\alpha_1\alpha_2\cdots\alpha_n|\hat a_l|\alpha_1\alpha_2\cdots\alpha_n\rangle\\&-\langle\alpha_1\alpha_2\cdots\alpha_n|\hat a^\dag_l|\alpha_1\alpha_2\cdots\alpha_n\rangle)\\
=&-i(\alpha_l-\alpha_l^*)=2{\rm Im}\alpha_l.
\end{array}$$
So $\bar{d}_{|\alpha\rangle\langle\alpha|}=(2{\rm Re}\alpha_1,2{\rm Im}\alpha_1,\ldots,2{\rm Re}\alpha_n,2{\rm Im}\alpha_n)$.

Next, for any $k,l$ with $1\le k,l\le n$, according to the definition of CM, one has

\begin{widetext}

	$$\begin{array}{rl}\nu_{2k-1,2l-1}=&\frac{1}{2}{\rm Tr}\big[|\alpha\rangle\langle\alpha|(\Delta\hat R_{2k-1}\Delta\hat R_{2l-1}+\Delta\hat R_{2l-1}\Delta\hat R_{2k-1})\big]\\
=&\frac{1}{2}{\rm Tr}\big[|\alpha\rangle\langle\alpha|((\hat Q_k-\hat d_{2k-1})(\hat Q_l-\hat d_{2l-1})
+(\hat Q_l-\hat d_{2l-1})(\hat Q_k-\hat d_{2k-1}))\big]\\
=&\frac{1}{2}{\rm Tr}\big[|\alpha\rangle\langle\alpha|((\hat Q_k-2{\rm Re}\alpha_k)(\hat Q_l-2{\rm Re}\alpha_l)+(\hat Q_l-2{\rm Re}\alpha_l)(\hat Q_k-2{\rm Re}\alpha_k))\big]\\
=&\frac{1}{2}{\rm Tr}\big[|\alpha\rangle\langle\alpha|(\hat Q_k\hat Q_l+\hat Q_l\hat Q_k)\big]-4{\rm Re}\alpha_k{\rm Re}\alpha_l\\
=&\frac{1}{2}{\rm Tr}\big[|\alpha\rangle\langle\alpha|(\hat a_k\hat a_l+\hat a_l\hat a_k+\hat a^\dag_k\hat a_l+\hat a_l\hat a^\dag_k+\hat a_k\hat a^\dag_l+\hat a^\dag_l\hat a_k+\hat a^\dag_k\hat a^\dag_l+\hat a^\dag_l\hat a^\dag_k)\big]-4{\rm Re}\alpha_k{\rm Re}\alpha_l\\
=&\left\{\begin{array}{ll}
{\rm Tr}\big[|\alpha\rangle\langle\alpha|(\hat a_k\hat a_l+\hat a^\dag_k\hat a_l+\hat a^\dag_l\hat a_k+\hat a^\dag_k\hat a^\dag_l)\big]-4{\rm Re}\alpha_k{\rm Re}\alpha_l,
& {\rm if}\ k\ne l,\\
{\rm Tr}\big[|\alpha\rangle\langle\alpha|((\hat a_k)^2+1+2\hat a^\dag_k\hat a_k+(\hat a^\dag_k)^2)\big]-4{\rm Re}\alpha_k{\rm Re}\alpha_l,
& {\rm if} \  k=l \end{array}\right.\\
=&\left\{\begin{array}{ll}
\alpha_k\alpha_l+\alpha_k^*\alpha_l+\alpha_l^*\alpha_k+\alpha_k^*\alpha_l^*-4{\rm Re}\alpha_k{\rm Re}\alpha_l, & {\rm if}\ k\ne l,\\
\alpha_k^2+1+2\alpha_k^*\alpha_k+(\alpha_k^*)^2-4{\rm Re}\alpha_k{\rm Re}\alpha_l,& {\rm if} \  k=l \end{array}\right.=\left\{\begin{array}{ll}
0, & {\rm if}\ k\ne l,\\
1,& {\rm if} \  k=l, \end{array}\right.
\end{array}$$
$$\begin{array}{rl}\nu_{2k-1,2l}=&\frac{1}{2}{\rm Tr}\big[|\alpha\rangle\langle\alpha|(\Delta\hat R_{2k-1}\Delta\hat R_{2l}+\Delta\hat R_{2l}\Delta\hat R_{2k-1})\big]\\
=&\frac{1}{2}{\rm Tr}\big[|\alpha\rangle\langle\alpha|((\hat Q_k-\hat d_{2k-1})(\hat P_l-\hat d_{2l})
+(\hat P_l-\hat d_{2l})(\hat Q_k-\hat d_{2k-1}))\big]\\
=&\frac{1}{2}{\rm Tr}\big[|\alpha\rangle\langle\alpha|((\hat Q_k-2{\rm Re}\alpha_k)(\hat P_l-2{\rm Im}\alpha_l)+(\hat P_l-2{\rm Im}\alpha_l)(\hat Q_k-2{\rm Re}\alpha_k))\big]\\
=&\frac{1}{2}{\rm Tr}\big[|\alpha\rangle\langle\alpha|(\hat Q_k\hat P_l+\hat P_l\hat Q_k)\big]-4{\rm Re}\alpha_k{\rm Im}\alpha_l\\
=&-i\frac{1}{2}{\rm Tr}\big[|\alpha\rangle\langle\alpha|(\hat a_k\hat a_l+\hat a_l\hat a_k+\hat a^\dag_k\hat a_l+\hat a_l\hat a^\dag_k-\hat a_k\hat a^\dag_l-\hat a^\dag_l\hat a_k-\hat a^\dag_k\hat a^\dag_l-\hat a^\dag_l\hat a^\dag_k)\big]-4{\rm Re}\alpha_k{\rm Im}\alpha_l\\
=&\left\{\begin{array}{ll}
-i{\rm Tr}\big[|\alpha\rangle\langle\alpha|(\hat a_k\hat a_l+\hat a^\dag_k\hat a_l-\hat a^\dag_l\hat a_k-\hat a^\dag_k\hat a^\dag_l)\big]-4{\rm Re}\alpha_k{\rm Im}\alpha_l,
& {\rm if}\ k\ne l,\\
-i{\rm Tr}\big[|\alpha\rangle\langle\alpha|((\hat a_k)^2-(\hat a^\dag_k)^2)\big]-4{\rm Re}\alpha_k{\rm Im}\alpha_k,
& {\rm if} \  k=l \end{array}\right.\\
=&\left\{\begin{array}{ll}
-i(\alpha_k\alpha_l+\alpha_k^*\alpha_l-\alpha_l^*\alpha_k-\alpha_k^*\alpha_l^*)-4{\rm Re}\alpha_k{\rm Im}\alpha_l, & {\rm if}\ k\ne l,\\
-i(\alpha_k^2-(\alpha_k^*)^2)-4{\rm Re}\alpha_k{\rm Im}\alpha_l,& {\rm if} \  k=l \end{array}\right.
=0,
\end{array}$$
and similarly,  $\nu_{2k,2l}=\left\{\begin{array}{rr}
	0, & {\rm if}\ k\ne l,\\
	1,& {\rm if} \  k=l. \end{array}\right.$ Since  $\nu\geq 0$, we have $\nu_{2l,2k-1}=\nu_{2k-1,2l}=0, 1\le k,l\le n$.
Hence $\nu_{|\alpha\rangle\langle\alpha|}=I_{2n}$.\hfill$\Box$

{\bf Proof of Theorem \ref{thm10}.}

(1) For $2\leq r<m$ and any Gaussian state $\rho=\rho_{1,2,\ldots,m}\in{\mathcal S}(H_1\otimes H_2\otimes\cdots\otimes H_m)$, we write $\rho=\rho(\nu_m,\bar{d}_m)$ and $\rho_K=\rho_K(\nu_r,\bar{d}_r)$, where $\rho_K={\rm Tr}_{K^c}(\rho)$ is the reduced state of $\rho$ to subsystem $K$, $\nu_m=(\nu_{kl})_{2m\times2m}$, $\nu_r=(\nu_{kl})_{2r\times2r}$, $\bar{d}_m=(d_1,d_2,\ldots,d_{2m})^{\rm T}$ and $\bar{d}_r=(d_1,d_2,\ldots,d_{2r})^{\rm T}$.
We have to check that
$${\mathcal I}_r^{G_{n_K}}(\rho_K)\leq {\mathcal I}_m^{G_{n_H}}(\rho).$$

Note that

	$$\begin{array}{rl}&P_{n}\nu_m P_{n}^{\rm T}=\\&\left( \begin{array}{ccc|ccc|ccc|ccc}
\nu_{11} & \cdots & \nu_{1,2r-1} & \nu_{1,2r+1}&\cdots &\nu_{1,2m-1}& \nu_{12} & \cdots & \nu_{1,2r} & \nu_{1,2r+2} & \cdots & \nu_{1,2m}\\
\nu_{31} & \cdots & \nu_{3,2r-1} & \nu_{3,2r+1}&\cdots &\nu_{3,2m-1}& \nu_{32} & \cdots & \nu_{132r} & \nu_{3,2r+2} & \cdots & \nu_{3,2m}\\
\vdots & \ddots & \vdots & \vdots & \ddots & \vdots &\vdots & \vdots & \ddots & \vdots & \ddots & \vdots\\
\nu_{2r-1,1} & \cdots & \nu_{2r-1,2r-1} & \nu_{2r-1,2r+1}&\cdots &\nu_{2r-1,2m-1}& \nu_{2r-1,2} & \cdots & \nu_{2r-1,2r} & \nu_{2r-1,2r+2} & \cdots & \nu_{2r-1,2m}\\
\hline
\nu_{2r+1,1} & \cdots & \nu_{2r+1,2r-1} & \nu_{2r+1,2r+1}&\cdots &\nu_{2r+1,2m-1}& \nu_{2r+1,2} & \cdots & \nu_{2r+1,2r} & \nu_{2r+1,2r+2} & \cdots & \nu_{2r+1,2m}\\
\vdots & \ddots & \vdots & \vdots & \ddots & \vdots &\vdots & \vdots & \ddots & \vdots & \ddots & \vdots\\
\nu_{2m-1,1} & \cdots & \nu_{2m-1,2r-1} & \nu_{2m-1,2r+1}&\cdots &\nu_{2m-1,2m-1}& \nu_{2m-1,2}& \cdots & \nu_{2m-1,2r} & \nu_{2m-1,2r+2} & \cdots & \nu_{2m-1,2m}\\
\hline
\nu_{21} & \cdots & \nu_{2,2r-1} & \nu_{2,2r+1}&\cdots &\nu_{2,2m-1}& \nu_{22} & \cdots & \nu_{2,2r} & \nu_{2,2r+2} & \cdots & \nu_{2,2m}\\
\nu_{41} & \cdots & \nu_{4,2r-1} & \nu_{4,2r+1}&\cdots &\nu_{4,2m-1}& \nu_{42} & \cdots & \nu_{4,2r} & \nu_{4,2r+2} & \cdots & \nu_{4,2m}\\
\vdots & \ddots & \vdots & \vdots & \ddots & \vdots &\vdots & \vdots & \ddots & \vdots & \ddots & \vdots\\
\nu_{2r,1} & \cdots & \nu_{2r,2r-1} & \nu_{2r,2r+1}&\cdots &\nu_{2r,2m-1}& \nu_{2r,2} & \cdots & \nu_{2r,2r} & \nu_{2r,2r+2} & \cdots & \nu_{2r,2m}\\
\hline
\nu_{2r+2,1} & \cdots & \nu_{2r+2,2r-1} & \nu_{2r+2,2r+1}&\cdots &\nu_{2r+2,2m-1}& \nu_{2r+2,2} & \cdots & \nu_{2r+2,2r} & \nu_{2r+2,2r+2} & \cdots & \nu_{2r+2,2m}\\
\vdots & \ddots & \vdots & \vdots & \ddots & \vdots &\vdots & \vdots & \ddots & \vdots & \ddots & \vdots\\
\nu_{2m,1} & \cdots & \nu_{2m,2r-1} & \nu_{2m,2r+1}&\cdots &\nu_{2m,2m-1}& \nu_{2m,2} & \cdots & \nu_{2m,2r} & \nu_{2m,2r+2} & \cdots & \nu_{2m,2m}\\
\end{array}\right)\\
=& \left(\begin{array}{cc|cc}
A_{11}  & A_{12} & A_{13}& A_{14}\\
A^{\rm T}_{12} & A_{22}& A_{23}& A_{24} \\
\hline
A^{\rm T}_{13} & A^{\rm T}_{23}& A_{33}& A_{34} \\
A^{\rm T}_{14} & A^{\rm T}_{24}& A^{\rm T}_{34}& A_{44} \\
\end{array}\right)=\left( \begin{array}{c|c}
B_{11}  & B_{12}\\
\hline
B_{12}^{\rm T} & B_{22}\\
\end{array}\right),\end{array}$$
and
$$\begin{array}{rl}P'P_{n}\nu_m P_{n}^{\rm T}P'^{\rm T}
	=&\left( \begin{array}{cc|cc}
A_{11}  & A_{13} & A_{12}& A_{14}\\
A^{\rm T}_{13} & A_{33}& A_{23}^{\rm T}& A_{34} \\
\hline
A_{12}^{\rm T} & A_{23}& A_{22}& A_{24} \\
A_{14}^{\rm T} & A_{34}^{\rm T}& A_{24}^{\rm T}& A_{44} \\
\end{array}\right)=\left( \begin{array}{c|c}
C_{11}  & C_{12}\\
\hline
C_{12}^{\rm T} & C_{22}\\
\end{array}\right),
\end{array}$$
where $\begin{array}{rl}P'=\left( \begin{array}{cccc}
I_{r} & 0 & 0 & 0\\
0 & 0 & I_{m-r} & 0 \\
0 & I_{r} & 0 & 0 \\
0 & 0 & 0 & I_{m-r} \\
\end{array}\right)\end{array}$ and $P'P'^{\rm T}=I_m$.
By Eq.\eqref{eq11}, one has
	\begin{equation}\label{eqA5}
\begin{array}{rl}{\mathcal I}_m^{G_{n_H}}(\rho)=&1-\dfrac{\det(\nu_m)}{\det(Q_{n_H}P_{n_H}\nu_m P_{n_H}^{\rm T}Q_{n_H}^{\rm T})\det(Q'_{n_H}P_{n_H}\nu_m P_{n_H}^{\rm T}Q_{n_H}'^{\rm T})}+h(\|Q_{n_H}'P_{n_H}\bar{d}_m\|_1)\\
=&1-\dfrac{\det(P'P_{n_H}\nu_m P_{n_H}^{\rm T}P'^{\rm T})}{\det(B_{11})\det(B_{22})}+h(\|Q_{n_H}'P_{n_H}\bar{d}_m\|_1)\\
=&1-\dfrac{\det(P'P_{n_H}\nu_m P_{n_H}^{\rm T}P'^{\rm T})}{\det(B_{11})\det(B_{22})}+h(|d_2|+|d_4|+\cdots+|d_{2r}|+\cdots+|d_{2m}|).
\end{array}\end{equation}
In addition,
$$\begin{array}{rl}P_{n_K}\nu_r P_{n_K}^{\rm T}=&\left( \begin{array}{cccc|cccc}
\nu_{11} & \nu_{13} & \cdots & \nu_{1,2r-1} & \nu_{12} & \nu_{14} & \cdots & \nu_{1,2r}\\
\nu_{31} & \nu_{33} & \cdots & \nu_{3,2r-1} & \nu_{32} & \nu_{34} & \cdots & \nu_{3,2r}\\
\vdots & \vdots & \ddots & \vdots &\vdots & \vdots & \ddots & \vdots \\
\nu_{2r-1,1} & \nu_{2r-1,3} & \cdots & \nu_{2r-1,2r-1} & \nu_{2r-1,2} & \nu_{2r-1,4} & \cdots & \nu_{2r-1,2r}\\
\hline
\nu_{21} & \nu_{23} & \cdots & \nu_{2,2r-1} & \nu_{22} & \nu_{24} & \cdots & \nu_{2,2r}\\
\nu_{41} & \nu_{43} & \cdots & \nu_{4,2r-1} & \nu_{42} & \nu_{44} & \cdots & \nu_{4,2r}\\
\vdots & \vdots & \ddots & \vdots & \vdots & \vdots & \ddots & \vdots \\
\nu_{2r,1} & \nu_{2r,3} & \cdots & \nu_{2r,2r-1} & \nu_{2r,2} & \nu_{2r,4} & \cdots & \nu_{2r,2r}\\
\end{array}\right)\\
=& \left(\begin{array}{cc}
A_{11}  & A_{13}\\
A^{\rm T}_{13} & A_{33}\\
\end{array}\right)=C_{11}.\end{array}$$
So, \begin{equation}\label{eqA6}
\begin{array}{rl}{\mathcal I}_r^{G_{n_K}}(\rho_K)=&1-\dfrac{\det(\nu_r)}{\det(Q_{n_K}P_{n_K}\nu_r P_{n_K}^{\rm T}Q_{n_K}^{\rm T})\det(Q'_{n_K}P_{n_K}\nu_r P_{n_K}^{\rm T}Q_{n_K}'^{\rm T})}+h(\|Q_{n_K}'P_{n_K}\bar{d}_r\|_1)\\
=&1-\dfrac{\det(C_{11})}{\det(A_{11})\det(A_{33})}+h(\|Q_{n_K}'P_{n_K}\bar{d}_r\|_1)\\
=&1-\dfrac{\det(C_{11})}{\det(A_{11})\det(A_{33})}+h(|d_2|+|d_4|+\cdots+|d_{2r}|).
\end{array}\end{equation}
Since $h(|d_2|+|d_4|+\cdots+|d_{2r}|)\le h(|d_2|+|d_4|+\cdots+|d_{2r}|+\cdots+|d_{2m}|)$, by Eqs.\eqref{eqA5}-\eqref{eqA6}, to prove ${\mathcal I}_r^{G_{n_K}}(\rho_K)\leq {\mathcal I}_m^{G_{n_H}}(\rho)$, we  only need to prove that
\begin{equation}\label{eqA7}
\dfrac{\det(C_{11})}{\det(A_{11})\det(A_{33})}\ge\dfrac{\det(P'P_{n_H}\nu_m P_{n_H}^{\rm T}P'^{\rm T})}{\det(B_{11})\det(B_{22})}.
\end{equation}
By using Lemma \ref{lem11}, we have $\det(P'P_{n_H}\nu_m P_{n_H}^{\rm T}P'^{\rm T})=\det(C_{11})\det(C_{22}-C_{12}^{\rm T} C_{11}^{-1}C_{12})$, $\det(B_{11})=\det(A_{11})\det(A_{22}-A_{12}^{\rm T} A_{11}^{-1}A_{12})$ and $\det(B_{22})=\det(A_{33})\det(A_{44}-A_{34}^{\rm T} A_{33}^{-1}A_{34})$. Hence the inequality in Eq.\eqref{eqA7} is equivalent to the following inequality:
\begin{equation}\label{eqA8}
\det(C_{22}-C_{12}^{\rm T} C_{11}^{-1}C_{12})\le \det(A_{22}-A_{12}^{\rm T} A_{11}^{-1}A_{12})\det(A_{44}-A_{34}^{\rm T} A_{33}^{-1}A_{34}).\end{equation}

For the positive definite matrix $C_{11}$, we have $$C_{11}^{-1}=\left(\begin{array}{cc}
(A_{11}-A_{13}A_{33}^{-1}A_{13}^{\rm T})^{-1} & -A_{11}^{-1}A_{13}(A_{33}-A_{13}^{\rm T}A_{11}^{-1}A_{13})^{-1}\\
-(A_{33}-A_{13}^{\rm T}A_{11}^{-1}A_{13})^{-1}A_{13}^{\rm T}A_{11}^{-1} & (A_{33}-A_{13}^{\rm T}A_{11}^{-1}A_{13})^{-1}\\
\end{array}\right),$$
and so  $$C_{22}-C_{12}^{\rm T} C_{11}^{-1}C_{12}=\left(\begin{array}{cc}
D_{11} & D_{12}\\
D_{12}^{\rm T} & D_{22}\\
\end{array}\right),$$
where
$$\begin{array}{rl}D_{11}=&A_{22}-A_{12}^{\rm T}(A_{11}-A_{13}A_{33}^{-1}A_{13}^{\rm T})^{-1}A_{12}+A_{23}(A_{33}-A_{13}^{\rm T}A_{11}^{-1}A_{13})^{-1}A_{13}^{\rm T}A_{11}^{-1}A_{12}\\
	&-A_{23}(A_{33}-A_{13}^{\rm T}A_{11}^{-1}A_{13})^{-1}A_{23}^{\rm T}+A_{12}^{\rm T}A_{11}^{-1}A_{13}(A_{33}-A_{13}^{\rm T}A_{11}^{-1}A_{13})^{-1}A_{23}^{\rm T},\end{array}$$
$$\begin{array}{rl}D_{12}=&A_{24}-A_{12}^{\rm T}(A_{11}-A_{13}A_{33}^{-1}A_{13}^{\rm T})^{-1}A_{14}+A_{23}(A_{33}-A_{13}^{\rm T}A_{11}^{-1}A_{13})^{-1}A_{13}^{\rm T}A_{11}^{-1}A_{14}\\&-A_{23}(A_{33}-A_{13}^{\rm T}A_{11}^{-1}A_{13})^{-1}A_{34}+A_{12}^{\rm T}A_{11}^{-1}A_{13}(A_{33}-A_{13}^{\rm T}A_{11}^{-1}A_{13})^{-1}A_{34}\end{array}$$ and
$$\begin{array}{rl}D_{22}=&A_{44}-A_{14}^{\rm T}(A_{11}-A_{13}A_{33}^{-1}A_{13}^{\rm T})^{-1}A_{14}+A_{34}^{\rm T}(A_{33}-A_{13}^{\rm T}A_{11}^{-1}A_{13})^{-1}A_{13}^{\rm T}A_{11}^{-1}A_{14}\\&-A_{34}^{\rm T}(A_{33}-A_{13}^{\rm T}A_{11}^{-1}A_{13})^{-1}A_{34}+A_{14}^{\rm T}A_{11}^{-1}A_{13}(A_{33}-A_{13}^{\rm T}A_{11}^{-1}A_{13})^{-1}A_{34}.\end{array}$$
Note that
$$\begin{array}{rl}&A_{12}^{\rm T}A_{11}^{-1}A_{12}\\
=& A_{12}^{\rm T} A_{11}^{-1}(A_{11}-A_{13}A_{33}^{-1}A_{13}^{\rm T})(A_{11}-A_{13}A_{33}^{-1}A_{13}^{\rm T})^{-1}A_{12}\\
=&A_{12}^{\rm T}(A_{11}-A_{13}A_{33}^{-1}A_{13}^{\rm T})^{-1}A_{12}-A_{12}^{\rm T}A_{11}^{-1}A_{13}A_{33}^{-1}A_{13}^{\rm T}(A_{11}-A_{13}A_{33}^{-1}A_{13}^{\rm T})^{-1}A_{12}\\
=&A_{12}^{\rm T}(A_{11}-A_{13}A_{33}^{-1}A_{13}^{\rm T})^{-1}A_{12}-A_{12}^{\rm T}A_{11}^{-1}A_{13}(A_{33}-A_{13}^{\rm T}A_{11}^{-1}A_{13})^{-1}A_{13}^{\rm T}A_{11}^{-1}A_{12}\end{array}$$
and
$$\begin{array}{rl}&A_{34}^{\rm T} A_{33}^{-1}A_{34}\\
	=& A_{34}^{\rm T} A_{33}^{-1}(A_{33}-A_{13}^{\rm T}A_{11}^{-1}A_{13})(A_{33}-A_{13}^{\rm T}A_{11}^{-1}A_{13})^{-1}A_{34}\\
	=& A_{34}^{\rm T}(A_{33}-A_{13}^{\rm T}A_{11}^{-1}A_{13})^{-1}A_{34}-A_{34}^{\rm T}A_{33}^{-1}A_{13}^{\rm T}A_{11}^{-1}A_{13}(A_{33}-A_{13}^{\rm T}A_{11}^{-1}A_{13})^{-1}A_{34}\\
	=& A_{34}^{\rm T}(A_{33}-A_{13}^{\rm T}A_{11}^{-1}A_{13})^{-1}A_{34}-A_{34}^{\rm T}A_{33}^{-1}A_{13}^{\rm T}
	(A_{11}-A_{13}A_{33}^{-1}A_{13}^{\rm T})^{-1}A_{13}A_{33}^{-1}A_{34}.
\end{array}$$
It follows that
$$\begin{array}{rl}&A_{22}-A_{12}^{\rm T} A_{11}^{-1}A_{12}-D_{11}\\=&A_{22}-A_{12}^{\rm T}(A_{11}-A_{13}A_{33}^{-1}A_{13}^{\rm T})^{-1}A_{12}+A_{12}^{\rm T}A_{11}^{-1}A_{13}(A_{33}-A_{13}^{\rm T}A_{11}^{-1}A_{13})^{-1}A_{13}^{\rm T}A_{11}^{-1}A_{12}\\
-&A_{22}+A_{12}^{\rm T}(A_{11}-A_{13}A_{33}^{-1}A_{13}^{\rm T})^{-1}A_{12}-A_{23}(A_{33}-A_{13}^{\rm T}A_{11}^{-1}A_{13})^{-1}A_{13}^{\rm T}A_{11}^{-1}A_{12}\\
+&A_{23}(A_{33}-A_{13}^{\rm T}A_{11}^{-1}A_{13})^{-1}A_{23}^{\rm T}-A_{12}^{\rm T}A_{11}^{-1}A_{13}(A_{33}-A_{13}^{\rm T}A_{11}^{-1}A_{13})^{-1}A_{23}^{\rm T}\\
=&A_{23}(A_{33}-A_{13}^{\rm T}A_{11}^{-1}A_{13})^{-1}A_{23}^{\rm T}-A_{23}(A_{33}-A_{13}^{\rm T}A_{11}^{-1}A_{13})^{-1}A_{13}^{\rm T}A_{11}^{-1}A_{12}\\
+&A_{12}^{\rm T}A_{11}^{-1}A_{13}(A_{33}-A_{13}^{\rm T}A_{11}^{-1}A_{13})^{-1}A_{13}^{\rm T}A_{11}^{-1}A_{12}-A_{12}^{\rm T}A_{11}^{-1}A_{13}(A_{33}-A_{13}^{\rm T}A_{11}^{-1}A_{13})^{-1}A_{23}^{\rm T}\\
=&A_{23}(A_{33}-A_{13}^{\rm T}A_{11}^{-1}A_{13})^{-1}(A_{23}^{\rm T}-A_{13}^{\rm T}A_{11}^{-1}A_{12})
-A_{12}^{\rm T}A_{11}^{-1}A_{13}(A_{33}-A_{13}^{\rm T}A_{11}^{-1}A_{13})^{-1}(A_{23}^{\rm T}-A_{13}^{\rm T}A_{11}^{-1}A_{12})\\
=&(A_{23}-A_{12}^{\rm T}A_{11}^{-1}A_{13})(A_{33}-A_{13}^{\rm T}A_{11}^{-1}A_{13})^{-1}(A_{23}-A_{12}^{\rm T}A_{11}^{-1}A_{13})^{\rm T}\ge0
\end{array}$$
and$$\begin{array}{rl}&A_{44}-A_{34}^{\rm T} A_{33}^{-1}A_{34}-D_{22}\\
=& A_{44}-A_{34}^{\rm T}(A_{33}-A_{13}^{\rm T}A_{11}^{-1}A_{13})^{-1}A_{34}+A_{34}^{\rm T}A_{33}^{-1}A_{13}^{\rm T}(A_{11}-A_{13}A_{33}^{-1}A_{13}^{\rm T})^{-1}A_{13}A_{33}^{-1}A_{34}\\
-& A_{44}+A_{14}^{\rm T}(A_{11}-A_{13}A_{33}^{-1}A_{13}^{\rm T})^{-1}A_{14}-A_{34}^{\rm T}(A_{33}-A_{13}^{\rm T}A_{11}^{-1}A_{13})^{-1}A_{13}^{\rm T}A_{11}^{-1}A_{14}\\
+& A_{34}^{\rm T}(A_{33}-A_{13}^{\rm T}A_{11}^{-1}A_{13})^{-1}A_{34}-A_{14}^{\rm T}A_{11}^{-1}A_{13}(A_{33}-A_{13}^{\rm T}A_{11}^{-1}A_{13})^{-1}A_{34}\\
=&A_{34}^{\rm T}A_{33}^{-1}A_{13}^{\rm T}(A_{11}-A_{13}A_{33}^{-1}A_{13}^{\rm T})^{-1}A_{13}A_{33}^{-1}A_{34}-A_{34}^{\rm T}(A_{33}-A_{13}^{\rm T}A_{11}^{-1}A_{13})^{-1}A_{13}^{\rm T}A_{11}^{-1}A_{14}\\
+&A_{14}^{\rm T}(A_{11}-A_{13}A_{33}^{-1}A_{13}^{\rm T})^{-1}A_{14}-A_{14}^{\rm T}A_{11}^{-1}A_{13}(A_{33}-A_{13}^{\rm T}A_{11}^{-1}A_{13})^{-1}A_{34}\\
=&A_{34}^{\rm T}A_{33}^{-1}A_{13}^{\rm T}(A_{11}-A_{13}A_{33}^{-1}A_{13}^{\rm T})^{-1}A_{13}A_{33}^{-1}A_{34}-A_{34}^{\rm T}A_{33}^{-1}A_{13}^{\rm T}(A_{11}-A_{13}A_{33}^{-1}A_{13}^{\rm T})^{-1}A_{14}\\
+&A_{14}^{\rm T}(A_{11}-A_{13}A_{33}^{-1}A_{13}^{\rm T})^{-1}A_{14}-A_{14}^{\rm T}
(A_{11}-A_{13}A_{33}^{-1}A_{13}^{\rm T})^{-1}A_{13}A_{33}^{-1}A_{34}\\
=& (A_{13}A_{33}^{-1}A_{34}-A_{14})^{\rm T}(A_{11}-A_{13}A_{33}^{-1}A_{13}^{\rm T})^{-1}(A_{13}A_{33}^{-1}A_{34}-A_{14})\ge0.
\end{array}$$
These indicate that
\begin{equation}\label{eqA9}
	D_{11}\le A_{22}-A_{12}^{\rm T} A_{11}^{-1}A_{12}\ \ {\rm and} \ \
	D_{22}\le A_{44}-A_{34}^{\rm T} A_{33}^{-1}A_{34}. \end{equation}
Combining Lemma \ref{lem12}(2) and Eq.\eqref{eqA9} yields that  Eq.\eqref{eqA8} holds.

(2) For any $m$-partite Gaussian state $\rho\in{\mathcal S}(H_1\otimes H_2\otimes \cdots\otimes H_m)$, regard also $\rho\in {\mathcal S}(H_{P_1}\otimes H_{P_2}\otimes\cdots\otimes H_{P_r})$ as an $r$-partite Gaussian state. Since ${\mathcal I}^{G_n}$ in Definition \ref{Df4} is defined through the mode instead of partition of quantum state,  it is obvious that  ${\mathcal I}_r^{G_{n_H}}(\rho)={\mathcal I}_m^{G_{n_H}}(\rho)$.

(3) For any $r$-partition $P=P_1|P_2|\ldots|P_r$ of $\{1,2,\ldots,m\}$, any $Q=Q_1|Q_2|\ldots|Q_r$ be obtained by $\emptyset\not= Q_j\subseteq P_j$, $j=1,2,\ldots,r$, and for any Gaussian state $\rho\in{\mathcal S}(H_1\otimes H_2\otimes\cdots\otimes H_m)$, regard $\rho\in{\mathcal S}(H_{P_1}\otimes H_{P_2}\otimes\cdots\otimes H_{P_r})$ and $\rho_Q\in{\mathcal S}(H_{Q_1}\otimes H_{Q_2}\otimes\cdots\otimes H_{Q_r})$ as $r$-partite Gaussian states. Since ${\mathcal I}^{G_{n}}$ is invariant under any permutation of subsystems, (1) clearly implies that ${\mathcal I}_r^{G_{n_{H_Q}}}(\rho_Q)\leq {\mathcal I}_r^{G_{n_H}}(\rho)$.

\hfill$\Box$

\end{widetext}


\end{document}